\documentclass[twocolumn, twocolappendix]{aastex631}

\newcommand{\D}{{\cal D}}
\newcommand{\J}{{\cal J}}

\usepackage{booktabs}
\usepackage{amsmath}
\usepackage{lipsum}
\usepackage{enumitem}
\usepackage[utf8]{inputenc} 
\usepackage[T1]{fontenc}
\usepackage{CJKutf8}

\newif\ifedits
\editsfalse
\ifedits
    \renewcommand{\edit}[1]{{\color{red} #1}}
    
\else
    \renewcommand{\edit}[1]{#1}
    
\fi

\defcitealias{Rose_2025}{R25a}
\defcitealias{CDM_Centrals_2025}{R25b}

\makeatletter
\let\frontmatter@title@above=\relax
\makeatother

\begin{document}

\shorttitle{The Impact of Physics and Halo-to-Halo Variations on Dark Matter Halos}
\shortauthors{Garcia et al.}

\title{The DREAMS Project: \\
Disentangling the Impact of Halo-to-Halo Variance and Baryonic Feedback on \\Milky Way Dark Matter Density Profiles}

\correspondingauthor{Alex M. Garcia}
\email{alexgarcia@virginia.edu}

\author[0000-0002-8111-9884]{Alex M. Garcia}
\affiliation{Department of Astronomy, University of Virginia, 530 McCormick Road, Charlottesville, VA 22904}
\affiliation{Virginia Institute for Theoretical Astronomy, University of Virginia, Charlottesville, VA 22904, USA}
\affiliation{The NSF-Simons AI Institute for Cosmic Origins, USA}

\author[0000-0002-2628-0237]{Jonah C. Rose}
\affiliation{Department of Physics, Princeton University, Princeton, NJ 08544, USA}
\affiliation{Center for Computational Astrophysics, Flatiron Institute, 162 5th Avenue, New York, NY 10010, USA}

\author[0000-0002-5653-0786]{Paul Torrey}
\affiliation{Department of Astronomy, University of Virginia, 530 McCormick Road, Charlottesville, VA 22904}
\affiliation{Virginia Institute for Theoretical Astronomy, University of Virginia, Charlottesville, VA 22904, USA}
\affiliation{The NSF-Simons AI Institute for Cosmic Origins, USA}

\author[0000-0003-1122-6606]{Andrea Caputo}
\affiliation{Theoretical Physics Department, CERN, 1211 Geneva 23, Switzerland}
\affiliation{Dipartimento di Fisica, “Sapienza” Università di Roma, Italy \& Sezione INFN Roma1, Piazzale Aldo Moro 5, 00185, Roma, Italy}
\affiliation{Department of Particle Physics and Astrophysics, Weizmann Institute of Science, Rehovot 7610001, Israel}

\author[0000-0002-8495-8659]{Mariangela Lisanti}
\affiliation{Department of Physics, Princeton University, Princeton, NJ 08544, USA}
\affiliation{Center for Computational Astrophysics, Flatiron Institute, 162 5th Avenue, New York, NY 10010, USA}

\author[0000-0002-6021-8760]{Andrew B. Pace}
\affiliation{Department of Astronomy, University of Virginia, 
530 McCormick Road, 
Charlottesville, VA 22904}

\author[0000-0003-2486-0681]{Hongwan Liu}
\affiliation{Physics Department, Boston University, Boston, MA 02215, USA}

\author[0000-0002-5560-8668]{Abdelaziz Hussein}
\affiliation{Department of Physics and Kavli Institute for Astrophysics and Space Research, Massachusetts Institute of Technology, Cambridge, MA 02139, USA}

\author{Haozhe Liu}
\affiliation{Department of Astronomy, University of Virginia,
530 McCormick Road, 
Charlottesville, VA 22904}

\author[0000-0002-4816-0455]{Francisco Villaescusa-Navarro}
\affiliation{Center for Computational Astrophysics, Flatiron Institute, 162 5th Avenue, New York, NY 10010, USA}


\author{John Barry} 
\affiliation{College of Science at Northeastern University, 360 Huntington Avenue, 115 Richards Hall, Boston, MA 02115}

\author[0009-0006-5740-5318]{Ilem Leisher} 
\affiliation{Grinnell College, 1115 8th Ave, Grinnell, IA 50112, USA}
\affiliation{Department of Physics, Massachusetts Institute of Technology, Cambridge, MA 02139, USA}
\affiliation{Kavli Institute for Astrophysics and Space Research, Massachusetts Institute of Technology, Cambridge, MA 02139, USA}

\author[0009-0008-6301-1539]{Belén Costanza} 
\affiliation{ Facultad de Ciencias Astron\'omicas y Geof\'isicas, Universidad Nacional de La Plata, Observatorio Astron\'omico, \\ Paseo del Bosque, B1900FWA La Plata, Argentina }
\affiliation{Consejo Nacional de Investigaciones Cient\'ificas y T\'ecnicas (CONICET), Rivadavia 1917, Buenos Aires, Argentina}

\author[0000-0002-0275-3001]{Jonathan Kho} 
\affiliation{Department of Astronomy, University of Virginia, 530 McCormick Road, Charlottesville, VA 22904}
\affiliation{Virginia Institute for Theoretical Astronomy, University of Virginia, Charlottesville, VA 22904, USA}
\affiliation{The NSF-Simons AI Institute for Cosmic Origins, USA}

\author[0009-0000-8180-9044]{Ethan Lilie} 
\affiliation{Department of Physics, Princeton University, Princeton, NJ 08544, USA}

\author[0000-0001-9592-4190]{Jiaxuan Li (\begin{CJK}{UTF8}{gbsn}李嘉轩\end{CJK}\!\!)} 
\affiliation{Department of Astrophysical Sciences, 4 Ivy Lane, Princeton University, Princeton, NJ 08540, USA}

\author[0009-0002-1233-2013]{Niusha Ahvazi} 
\affiliation{Department of Astronomy, University of Virginia, 530 McCormick Road, Charlottesville, VA 22904}
\affiliation{Virginia Institute for Theoretical Astronomy, University of Virginia, Charlottesville, VA 22904, USA}
\affiliation{The NSF-Simons AI Institute for Cosmic Origins, USA}

\author[0000-0002-7080-2864]{Aklant Bhowmick} 
\affiliation{Department of Astronomy, University of Virginia, 530 McCormick Road, Charlottesville, VA 22904}
\affiliation{Virginia Institute for Theoretical Astronomy, University of Virginia, Charlottesville, VA 22904, USA}
\affiliation{The NSF-Simons AI Institute for Cosmic Origins, USA}

\author[0000-0001-6189-8457]{Tri Nguyen} 
\affiliation{Center for Interdisciplinary Exploration and Research in Astrophysics, Northwestern University, 1800 Sherman Ave, Evanston, IL 60201}
\affiliation{NSF-Simons AI Institute for the Sky, 172 E. Chestnut St., Chicago, IL 60611, USA}

\author[0000-0002-7968-2088]{Stephanie O'Neil} 
\affiliation{Department of Physics \& Astronomy, University of Pennsylvania, Philadelphia, PA 19104, USA}
\affiliation{Department of Physics, Princeton University, Princeton, NJ 08544, USA}

\author[0000-0002-4669-9967]{Xiaowei Ou} 
\affiliation{Department of Astronomy, University of Virginia, 530 McCormick Road, Charlottesville, VA 22904}
\affiliation{Virginia Institute for Theoretical Astronomy, University of Virginia, Charlottesville, VA 22904, USA}
\affiliation{The NSF-Simons AI Institute for Cosmic Origins, USA}

\author[0000-0002-6196-823X]{Xuejian Shen} 
\affiliation{Department of Physics and Kavli Institute for Astrophysics and Space Research, Massachusetts Institute of Technology, Cambridge, MA 02139, USA}

\author[0000-0003-0777-4618]{Arya Farahi} 
\affiliation{Departments of Statistics and Data Sciences, University of Texas at Austin, Austin, TX 78757, USA}
\affiliation{The NSF-Simons AI Institute for Cosmic Origins, USA}

\author[0000-0002-3204-1742]{Nitya Kallivayalil} 
\affiliation{Department of Astronomy, University of Virginia, 530 McCormick Road, Charlottesville, VA 22904}
\affiliation{The NSF-Simons AI Institute for Cosmic Origins, USA}

\author[0000-0003-2806-1414]{Lina Necib} 
\affiliation{Department of Physics and Kavli Institute for Astrophysics and Space Research, Massachusetts Institute of Technology, Cambridge, MA 02139, USA}
\affiliation{The NSF AI Institute for Artificial Intelligence and Fundamental Interactions, Cambridge, MA 02139, USA}

\author[0000-0001-8593-7692]{Mark Vogelsberger} 
\affiliation{Department of Physics and Kavli Institute for Astrophysics and Space Research, Massachusetts Institute of Technology, Cambridge, MA 02139, USA}
\affiliation{The NSF AI Institute for Artificial Intelligence and Fundamental Interactions, Cambridge, MA 02139, USA}
\affil{Fachbereich Physik, Philipps Universit\"at Marburg, D-35032 Marburg, Germany}
\begin{abstract}
In this work, we utilize a new suite of Milky Way-mass halos from the DREAMS Project, simulated with Cold Dark Matter~(CDM), to quantify the influence of baryon feedback and intrinsic halo-to-halo variance on dark matter density profiles.
Our suite of 1024 halos 
varies over supernova and black hole feedback parameters from the IllustrisTNG model, as well as variations in two cosmological parameters.
We find that, \edit{for the DREAMS parameter variations}, Milky Way-mass dark matter density profiles in the IllustrisTNG model are largely insensitive to astrophysics and cosmology variations, with the dominant source of scatter instead arising from halo-to-halo variance.
However, most of the (comparatively minor) feedback-driven variations come from the changes to supernova prescriptions.
By comparing to dark matter-only simulations, we find that the strongest supernova wind energies are so effective at preventing galaxy formation that the halos are nearly entirely collisionless dark matter.
Finally, regardless of physics variation, all the DREAMS halos are roughly consistent with a halo contracting adiabatically from the presence of baryons, unlike models that have bursty stellar feedback.
This work represents a step toward assessing the \edit{uncertainty in} Milky Way dark matter profiles, with direct implications for dark matter searches where systematic uncertainty in the density profile remains a major challenge.
\end{abstract}

\keywords{Hydrodynamical simulations~(767)	--- Dark matter distribution~(356) --- Cold dark matter~(265)}

\section{Introduction} \label{sec:intro}

Baryons make up only a small fraction of the total mass budget of the Universe (\citeauthor{Planck_2016} \citeyear{Planck_2016}), yet they play a central role in driving galaxy evolution.
Unlike the dark matter component, which is assumed to be cold and collisionless in the standard Cold Dark Matter~(CDM) model, the baryonic component experiences \edit{hydrodynamical} forces and energy dissipation.
These interactions ultimately facilitate the condensation of gas at the centers of halos and the formation of the galaxies we observe \citep{Fall_1980,Blumenthal_1986}.
Once a galaxy is formed, baryons can impart feedback in their environments via various processes including supernovae, stellar winds, and active galactic nuclei~(AGN) activity  \citep[e.g.,][]{Silk_Dekel_1986,Larson_1974,Somerville_Dave_2015}. 
Through interactions, dissipation, and feedback, baryons drive time-varying changes in the gravitational potential that shape galaxy evolution.

The role that baryons play can be clearly seen in the distribution of matter within a halo.
Early dark matter-only~(DMO) simulations showed that halos ubiquitously form a two-component power-law density profile known as the \citeauthor*{NFW_1997} (\citeyear{NFW_1997}; NFW) profile, such that
\begin{equation}
    \label{eqn:NFW}
    \rho(r) = \cfrac{\rho_s}{\left(\frac{r}{r_s}\right)\left[1+\left(\frac{r}{r_s}\right)\right]^2}~,
\end{equation}
where $\rho_s$ is the scale density and $r_s$ is the scale radius of the profile.
Interior to the scale radius, the NFW profile has a power-law index of~$-1$ which then transitions beyond the scale radius to a power law with an index of~$-3$.
From an observational perspective, measurements of galactic rotation curves provide constraints for the distribution of the mass within a halo \citep[][]{Rubin_1980,Persic_1996,Sofue_2001,Huang_2016,Zhang_2024}.
These observations show that, while massive halos (e.g., clusters, $\sim10^{14}~{\rm M}_\odot$) tend to be ``cuspy'' (roughly consistent with the NFW inner power-law index of $\sim-1$; e.g., \citeauthor{Newman_2013}~\citeyear{Newman_2013}), observed dwarf galaxies ($\sim10^{9-10}~{\rm M}_\odot$) can have a diversity of cuspy and ``cored'' (inner power law $> -1$) density profiles~\citep{deBlok_2002,deBlok_2008,Walker_2011,Oman_2015}.
The striking diversity of density profiles contrasts with the relatively uniform predictions of DMO simulations, a discrepancy known as the diversity problem \citep[e.g.,][]{Oman_2015,Bullock_2017}, a generalization of the core-cusp problem \citep{deBlock_2010}.

Simulations which self-consistently model the baryonic component provide one possible resolution to these tensions \citep{Sales_2022}.
A cored inner density can be achieved using sufficiently ``bursty'' supernova feedback~\citep{Chan_2015,Lazar_2020,Mostow_2024}.
Repetitively cycling between slow/smooth inflows followed by fast/impulsive outflows creates central potential fluctuations that disrupt the orbits of dark matter particles and lead to decreased central densities 
\citep[][although this effect is mass-dependent with it being most pronounced in dwarf galaxies]{Governato_2010, Pontzen_2012, Onorbe_2015, Jahn_2023}.
However, even at fixed physics prescriptions, the diversity of halo assembly histories introduces substantial halo-to-halo variance in the resulting dark matter profiles \cite[e.g.,][]{Abadi_2010,Duffy_2010,DiCintio_2014b,Chan_2015,Fitts_2017,Lazar_2020,Farahi_2022profile}.
Disentangling this intrinsic variance from systematic shifts due to baryonic physics is therefore critical for interpreting both simulations and observations.
Moreover, the level to which feedback is generically bursty and its evolutionary consequences across different galaxy populations remains uncertain \citep[see, e.g., work by][]{Hartley_2016,Bhagwat_2024,Garcia_2023,Garcia_2024b,Garcia_2024,Garcia_2025,Garcia_2025a,Garcia_2025c}.

Current cosmological simulations of galaxy evolution follow several prescriptions for modeling the physics of baryons \citep[see, e.g.,][for reviews]{Vogelsberger_2020,Crain_2023,Feldmann_2025}.
These prescriptions often rely on ``subgrid'' models for the numerical implementations of unresolved physics.
For example, the interstellar medium (ISM) is often treated with an effective equation of state that sets the behavior of dense ($n_{\rm H}\gtrsim 0.1~{\rm cm}^{-3}$) star-forming gas \citep{Springel_Hernquist_2003,Schaye_DallaVechhia_2008}.
Subgrid models require assumptions and oftentimes contain tunable parameters within the model.
As a result, different simulation models make different predictions for the role of baryons and their associated feedback---even with relatively similar implementations \citep[e.g.,][]{Chua_2019,Chua_2022,Garcia_2024b,Garcia_2025,Garcia_2025a}.
Some work has been done to quantify the variation in model predictions based on the input physics \citep[e.g.,][]{Duffy_2010,Torrey_2014,Pillepich_2018,Chua_2019,Chua_2022,Font_2020,Anbajagane_Baryonic_imprints,Kugel_2023}, but the number of model variations is usually few (on the order of $\sim10$s of model variations).
Recent simulation efforts such as CAMELS~\citep[][]{CAMELS_2021,Ni_2023} and DREAMS~\citep[][hereafter \citetalias{Rose_2025}]{Rose_2025} have begun to more systematically explore the parameter space of well-tested simulation models via thousands of model variations.
These efforts allow for quantification of parameter uncertainties  while also providing large enough samples to begin quantifying the intrinsic halo-to-halo variance.

One model that departs sharply from the standard approach is FIRE~\citep[][]{FIRE,FIRE2}, which explicitly resolves the multiphase interstellar medium.
While FIRE has its own subgrid assumptions and parameters, the treatment of the star-forming ISM is a marked improvement compared to the standard approach of an effective equation of state.
One feature of the FIRE model is the production of strong, time-variable stellar feedback-driven winds \citep[i.e., bursty feedback;][]{Muratov_2015}, whereas models that adopt effective equation of state ISM treatment yield smoother, less variable stellar feedback.
In the absence of this bursty feedback, the evolution of a dark matter halo is governed by the response of the dark matter to the baryons cooling to the center.
This contraction can be modeled via the conservation of action integrals \citep[e.g.,][]{Eggen_1962,Blumenthal_1986,Gnedin_2004,Cautun_2020}: as the potential deepens due to baryon mass accumulation, the radius of the dark matter orbits decreases causing the total density in the center to increase.
Feedback can partially offset or even reverse this contraction, but the degree depends sensitively on feedback \citep{Brooks_2014,Lovell_2018,Hussein_2025}.
Indeed, \cite{Hussein_2025} show that FIRE-2 galaxies deviate much more from \edit{simple} adiabatic contraction \edit{models} than simulations with equation of state ISM prescriptions, though it remains unclear whether this owes specifically to burstiness or just the overall feedback strength.

In this work, we employ the new DREAMS CDM suite of Milky Way-mass halos to quantify the role of variations to the IllustrisTNG physics model on the dark matter density profiles of halos.
The novelty of this DREAMS suite is that it contains 1024 Milky Way-mass halos, a factor of five more than the TNG volume at comparable resolution (TNG50; \citeauthor{Pillepich_2019}~\citeyear{Pillepich_2019}), as well as systematic variations in our baryonic feedback prescriptions.
Crucially, the large number of halos also allows us to quantify the contribution of halo-to-halo variance and to compare its relative importance to the impact of physics variations.

The rest of this paper is organized as follows.
In Section~\ref{sec:methods}, we introduce the DREAMS simulations, their parameter variations, and halo selection. We then outline our method for reconstructing dark matter density fields, describe the neural network emulators used in this work, and discuss a scheme to account for how well our halos match observed galaxy scaling relations.
In Section~\ref{sec:results}, we discuss the dark matter density profiles and fit them with an analytic function.
In Section~\ref{sec:discussion}, we compare the hydrodynamic simulations to DMO simulations, quantify the contraction the halos must have undergone due to the presence of baryons, and contrast our results to those from the bursty feedback FIRE model.
In Section~\ref{sec:conclusions}, we state our conclusions.


\section{Methods} \label{sec:methods}

In this Section, we 
describe the DREAMS CDM suite~(\S\ref{subsec:DREAMS}), describe how we reconstruct and fit the dark matter density profiles~(\S\ref{subsec:density_reconstruction}),  introduce our neural network emulation scheme~(\S\ref{subsec:emulator}), and summarize the~\cite[][hereafter \citetalias{CDM_Centrals_2025}]{CDM_Centrals_2025} halo weighting scheme we apply to our sample~(\S\ref{subsec:weighting}).

\subsection{DREAMS Simulations}
\label{subsec:DREAMS}

\begin{table*}
    \centering
    \begin{tabular}{p{0.1\linewidth}p{0.45\linewidth}p{0.2\linewidth}p{0.15\linewidth}}
        \toprule
         \textbf{Parameter} & \textbf{Brief Description} & \textbf{Fiducial TNG Value} & \textbf{Variations} \\\midrule
         $\bar{e}_w$ & SN Wind Energy & 3.6 & $[0.25$--4]$ \times3.6$ \\
         $\kappa_w$ & SN Wind Speed & 7.4 & $[0.5$--$2]\times7.4$\\
         $\epsilon_{f,\,\mathrm{high}}$ & AGN Feedback Strength & 0.1 & $[0.25$--$4]\times0.1$\\
         $\Omega_{\rm M}$ & Matter Density of Universe & 0.31 & $[0.274$--$0.354]$ \\
         $\sigma_8$ & Amplitude of Matter Clustering on 8 Mpc Scales & 0.8159 & $[0.780$--$0.888]$
         \\\bottomrule
    \end{tabular}
    \caption{{\bf Astrophysical and Cosmological Parameter Variations.}
    The astrophysical ($\bar{e}_w$, $\kappa_w$, $\epsilon_{f,\,\mathrm{high}}$) and cosmological ($\Omega_{\rm M}$, $\sigma_8$) parameters varied in the CDM DREAMS Milky Way-mass zoom-in simulations.
    We list the fiducial TNG value for each parameter, the variations used in these simulations (corresponding to a factor of $4$ for $\bar{e}_w$ and $\epsilon_{f,\,\mathrm{high}}$, a factor of $2$ for $\kappa_w$, and the two-standard-deviation uncertainty on \protect\citeauthor{Planck_2014}~\protect\citeyear{Planck_2014} for $\Omega_{\rm M}$ and $\sigma_8$).
    We sample the unique set of parameters for each simulation according to a \protect\cite{sobol} sequence.
    Finally, we note that $\bar{e}_w$, $\kappa_w$, and $\epsilon_{f,\,\mathrm{high}}$ are sampled in logarithmic space, whereas $\Omega_{\rm M}$ and $\sigma_8$ are sampled in linear space.
    Detailed justifications for these parameter ranges can be found in \protect\citetalias{Rose_2025}.
    }
    \label{tab:dreams_parameters}
\end{table*}

This work uses data products from the DREAMS suites of Milky Way-mass halos with cold dark matter \citepalias{Rose_2025,CDM_Centrals_2025}.\ignorespaces
\footnote{\href{https://www.dreams-project.org/}{dreams-project.org}
}
Each target halo falls within the mass range $5\times10^{11}~{\rm M}_\odot < M_{\rm Halo} < 2.0\times10^{12}~{\rm M}_\odot$, roughly corresponding to the mass of the Milky Way halo \citep{Callingham_2019,Wang_2020_MW}, where $M_{\rm Halo}$ is the sum of all mass within $R_{200}$ (\citeauthor{Springel_2005} \citeyear{Springel_2005}).
There are 2048 simulations in the DREAMS Milky Way CDM suite: 1024 DMO simulations as well as a corresponding pair of 1024 simulations with baryon physics included.
Each pair has the same initial conditions, such that the same halo is simulated twice.

The hydrodynamic simulations are based on the IllustrisTNG (hereafter simply TNG) physics model~\citep{Marinacci_2018,Naiman_2018,Nelson_2018,Pillepich_2018b,Springel_2018,Pillepich_2019, Nelson_2019a, Nelson_2019b}.
TNG is built on the moving Voronoi mesh code {\sc arepo} \citep{Springel_2010}.
The TNG model implements a wide range of astrophysical processes, including (but not limited to) feedback from stars, the growth of supermassive black holes, feedback from AGN, cosmological expansion of the universe, gravity, dark matter, and galactic winds; we refer the reader to~\citet{Pillepich_2018} for a complete description of the TNG model.

The DREAMS halos are selected using an iterative zoom-in process 
(described in detail in \citetalias{Rose_2025}, \citetalias{CDM_Centrals_2025}).
Briefly, the halos are selected initially from a low-resolution $(100~{\rm Mpc}/h)^3$ DMO simulation and are then resimulated with DMO zooms.
The target halos of the intermediate zooms are selected to have masses of $5\times10^{11}~{\rm M}_\odot < M_{\rm Halo} < 2.0\times10^{12}~{\rm M}_\odot$
and no massive~($M_{\rm Halo} > 10^{12}~{\rm M}_\odot$) galaxy within $1$~Mpc at $z~\!\!=~\!\!0$.
These zoom-in systems are then resimulated with hydrodynamics and baryonic feedback turned on and increasing the resolution of all particles within $5R_{200}$.

The key advantage of the DREAMS simulations is that they contain systematic variations over the fiducial TNG astrophysical ($\bar{e}_w$, $\kappa_w$, and $\epsilon_{f,\,\mathrm{high}}$) and cosmological parameters ($\Omega_{\rm M}$ and $\sigma_8$), as well as a range of initial conditions that sample rare environments and formation histories.
The simulation parameters are varied according to a \cite{sobol} sequence where each simulation represents its own unique realization of TNG physics and cosmology (i.e., SB5 or ``Sobol 5'' in the parlance of the CAMELS simulations; \citeauthor{CAMELS_2021} \citeyear{CAMELS_2021}).
Detailed motivation for the parameter ranges can be found in \citetalias{Rose_2025}~(their Section~2.1).
In the following two subsections, we define each of the varied parameters as well as their ranges, as summarized in Table~\ref{tab:dreams_parameters}.
Aside from the two cosmological parameters explicitly varied, the rest of the cosmology in DREAMS is fully consistent with \cite{Planck_2016}: $H_0=100h~{\rm km\, s^{-1}\, Mpc^{-1}}$ where $h=0.6909$, and $\Omega_{b}=0.046$.

The dark matter mass resolution of both the hydro and DMO simulations is implicitly varied by the variations in $\Omega_{\rm M}$, the matter density of the universe.
The dark matter mass resolution in the hydro simulations is $1.8\times(\Omega_{\rm M}/0.314)\times10^{6}~{\rm M}_\odot$, while the baryon mass resolution is $2.8\times10^5~{\rm M}_\odot$ (approximately a factor of four coarser resolution than the TNG50 suite; \citeauthor{Pillepich_2019} \citeyear{Pillepich_2019}).
Importantly, the DMO simulations are generated with a baryon density of $\Omega_{b}=0.046$ to preserve baryon acoustic oscillations, but the baryon particles are treated as collisionless.
In other words, the dark matter particles in the DMO simulations are more massive by a factor of $(\Omega_{b} - \Omega_{\rm M})/\Omega_{\rm M}$, meaning that DMO simulations have dark matter mass resolutions of $2\times(\Omega_{\rm M}/0.314)\times10^6~{\rm M}_\odot$.
All other numerical parameters (e.g., softening lengths of $0.441$ kpc) are the same between the two sets of simulations.

\subsubsection{Parameter Variations}
\label{subsec:astro_variations}

There are three astrophysical parameters---$\bar{e}_w$, $\kappa_w$, and $\epsilon_{f,\,\mathrm{high}}$---that are varied in every DREAMS simulation. 
Both $\bar{e}_w$ and $\kappa_w$ relate to the feedback from stars in the TNG model, whereas $\epsilon_{f,\,\mathrm{high}}$ relates to the feedback from supermassive black holes.

Given the limited mass resolution of TNG, feedback from stars is necessarily treated on a `subgrid' level.
This subgrid prescription accounts for the ejection of winds from unresolved supernovae explosions.
This feedback is primarily governed in the TNG model by the mass-loading factor of winds generated by the supernovae.
The mass-loading factors in TNG are given by the expression
\begin{equation}
    \label{eqn:mass_loading}
    \eta_w = \frac{2}{v_w^2}e_w\left(1-\tau_w\right)~.
\end{equation}
There are three free parameters in Equation~\ref{eqn:mass_loading} that decide the strength of feedback: specific energy of the winds~($e_w$), the wind velocity~($v_w$), and the fraction of thermal energy that is released~($\tau_w$).
The DREAMS simulations vary both the energy and speed of the winds while keeping the fraction of thermal energy constant to its fiducial TNG value~($\tau_w=0.1$).

The specific energy is defined such that
\begin{align}
    e_w =&~ \bar{e}_w\cdot f(Z)\cdot N_{\rm SN}\,\left[10^{51}\frac{\rm erg}{{\rm M}_\odot}\right]~,
\end{align}
where $\bar{e}_w$ is a dimensionless scaling on the energy per supernova, $f(Z)$ is a function that reduces the available energy when the metallicity~($Z$) is above a reference value, and $N_{\rm SN}$ is the number of Type II supernovae.
$N_{\rm SN}$ depends both on the shape and mass limits of the assumed initial mass function~(IMF): both TNG and DREAMS assume a \cite{Chabrier_2003} IMF with a minimum core-collapse supernova mass of $8~{\rm M}_\odot$.
The fiducial TNG model assumes $\bar{e}_w=3.6$.
The DREAMS simulations vary over $\bar{e}_w$ in logarithmic space from $0.9$ to $14.4$.\footnote{\ignorespaces
The parameter variations are performed in log space for $\bar{e}_w$, $\kappa_w$, and $\epsilon_{f,\,\mathrm{high}}$ because these are all multiplicative factors.
Varying them in log space ensures that the sampling is symmetric in terms of fractional changes rather than absolute values. 
In practice, this is especially useful since these amplitudes can span several orders of magnitude: a log scaling guarantees that, e.g., doubling and halving a parameter are treated as equally significant variations.
}

The speed of stellar winds is defined such that
\begin{equation}
    v_w = \operatorname{max}\left[\kappa_w\,\sigma_{\scriptscriptstyle{\rm DM}}\left(\frac{H_0}{H(z)}\right)^{1/3}, v_{w,\,{\rm min}}\right]~,
\end{equation}
where $\kappa_w$ is a dimensionless normalization factor, $H(z)$ is the Hubble parameter at redshift $z$, $v_{w,\,{\rm min}}$ is the minimum wind speed (set to $350~{\rm km\,s^{-1}}$), and $\sigma_{\scriptscriptstyle{\rm DM}}$ is the velocity dispersion of the 64 nearest dark matter particles.
The fiducial value of $\kappa_w$ in TNG is 7.4; in DREAMS, it is varied logarithmically from $3.7$ to $14.8$.
The imposed floor on the wind speed can potentially diminish the impact of varying $\kappa_w$, particularly in low-mass satellites and at high redshifts.
Indeed, the wind speed floor in the fiducial TNG model increases the efficacy of feedback at early times \citep{Pillepich_2018}.

Note that $\bar{e}_w$ and $\kappa_w$ enter in the mass-loading factor of Equation~\ref{eqn:mass_loading} differently.
The energy injection rate can be quantified via
\begin{equation}
    \dot{E}_{\rm wind} \propto \edit{\eta_w} v_w^2~.
\end{equation}
The impact of $\bar{e}_w$ is clear: increasing $\bar{e}_w$ raises the mass-loading factor, which increases the energy injection rate.
Increases to $\kappa_w$, on the other hand, have a more subtle effect.
By increasing $\kappa_w$, the wind speed increases, however, the mass-loading factor is proportionally decreased~(since $\edit{\eta_w}\propto v_w^{-2}$) keeping the energy injection rate fixed.
Separate from the energy injection rate, the ``efficacy'' of feedback is also important. 
That is, whether launching less material faster has a stronger impact on the system than launching more material slower.
Here we define ``efficacy'' in terms of the galaxy-scale consequences of feedback, such as reduced stellar mass growth and stronger baryon removal, instead of the instantaneous energy injection rate (which is fixed in the case of $\kappa_w$ variations).
For the range of values probed in this work, increasing $\kappa_w$ increases the ``efficacy''---by our definition---of feedback-driven winds~(i.e., in this part of the TNG model parameter space it is more effective to drive slightly less material at slightly faster speeds).
We show this effect explicitly in Section~\ref{subsubsec:mass_growth}.

Feedback from AGN in the TNG model is implemented in one of two channels based on the accretion rate of the central supermassive black hole (see~\citeauthor{Weinberger_2018} \citeyear{Weinberger_2018}, for a complete description of the AGN and black hole model in TNG).
Low accretion rates correspond to a `kinetic mode' of feedback where winds are driven in a pulsed and directed fashion, whereas high accretion rates describe a `thermal mode' of feedback where thermal energy is continuously dumped into the surroundings.
The thermal mode of feedback  dominates for TNG galaxies with stellar masses $\lesssim10^{10.5}~{\rm M}_\odot$, while galaxies with larger stellar masses are dominated by kinetic-mode feedback~\citep{Weinberger_2017}.
Since the Milky Way hosts in  DREAMS spend most---if not all---of their lives below this stellar mass, only the thermal feedback mode is varied in this suite.
Specifically, the normalization of the continuous thermal-feedback energy is varied, given by
\begin{equation}
    \label{eqn:agn}
    \dot{E}_{\rm AGN} = \epsilon_r\epsilon_{f,\,\mathrm{high}}\dot{M}_{\rm BH}c^2~,
\end{equation}
where $\epsilon_r$ is the radiative efficiency, $\epsilon_{f,\,\mathrm{high}}$ is the fraction of energy that is transferred to the nearby gas, and $\dot{M}_{\rm BH}$ is the accretion rate onto the black hole.
The fiducial $\epsilon_{f,\,\mathrm{high}}$ parameter is $0.1$ in TNG and is varied logarithmically from $0.025$ to $0.4$ in DREAMS.

DREAMS takes a wide prior for the cosmology variations consistent with $2\sigma$ variations from \cite{Planck_2014} for $\sigma_8$~(ranging from 0.780 to 0.888) and $\Omega_{\rm M}$~(ranging from 0.274 to 0.354).
This range contains the fiducial TNG values of $\Omega_{\rm M}=0.31$ and $\sigma_8=0.8159$ \citep[][taken from \citeauthor{Planck_2016} \citeyear{Planck_2016}]{Pillepich_2018} and is intentionally kept larger than current observational uncertainties to minimize prior effects on our results.
The two cosmological parameter variations are sampled linearly in DREAMS.

\begin{figure*}
    \centering
    \includegraphics[width=0.98\linewidth]{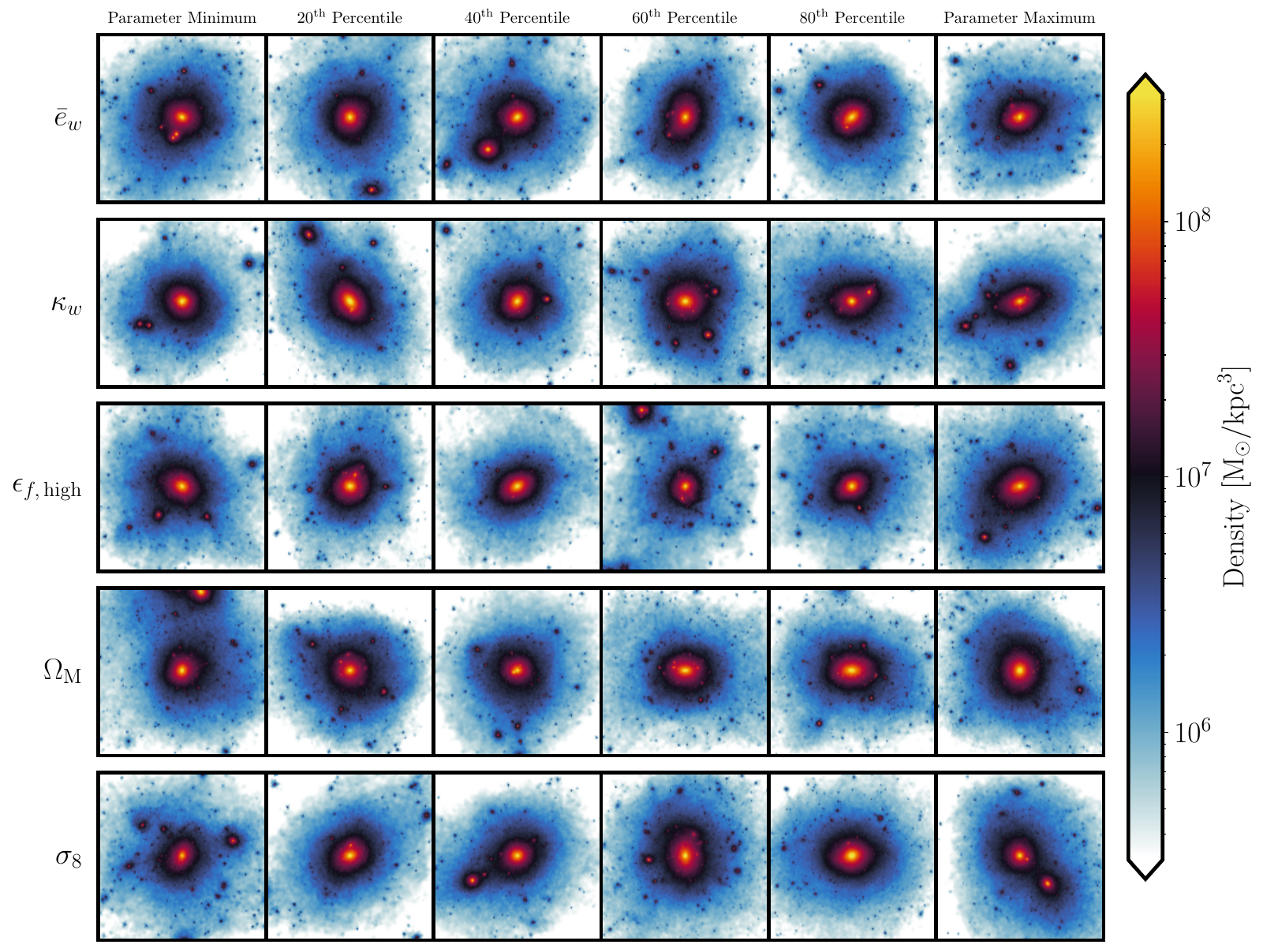}
    \caption{{\bf Milky Way Mass Dark Matter Halos from the SB5 DREAMS CDM Suite.}
    Projections of the dark matter density of a small fraction ($3\%$) of DREAMS halos.
    Each row contains variations of the five DREAMS parameters ($\bar{e}_w$, $\kappa_w$, $\epsilon_{f,\,\mathrm{high}}$, $\Omega_{\rm M}$, and $\sigma_8$, top-to-bottom, respectively; see Table~\ref{tab:dreams_parameters} and Section~\ref{subsec:astro_variations}).
    Each column represents the minimum, $20^{\rm th}$ percentile, $40^{\rm th}$ percentile, $60^{\rm th}$ percentile, $80^{\rm th}$ percentile, and maximum variation for each parameter, noting that in each row the other DREAMS parameters are also co-varied.
    }
    \label{fig:pretty_picture}
\end{figure*}

We show a small sample~($3\%$) of DREAMS halos in Figure~\ref{fig:pretty_picture}, which shows the dark matter density within $R_{200}$.
Each row of Figure~\ref{fig:pretty_picture} represents variations of the five DREAMS parameters with the columns ranging from the minimum value~(left-most) we probe to the maximum value~(right-most) noting that the simulations are not strictly one-parameter variations~(see above discussion).

\subsection{Reconstruction of Dark Matter Density}
\label{subsec:density_reconstruction}

The simulation dark matter is modeled as a collection of (large) discrete particles in TNG~(see Section~\ref{subsec:DREAMS} for mass resolution).
Since the goal of this paper is to study the dark matter profiles, we need to ``reconstruct'' a continuous density from these discrete particles.
To obtain the density at each point in space, we distribute the mass of a single dark matter particle according to a smooth particle hydrodynamics kernel weighting scheme \citep[e.g.,][]{Monaghan_1992,Springel_2010b}.
The density at a location in space, $s$, is computed as
\begin{equation}
    \rho_s = \sum_n^{N_{\rm ngb}} m_n W(r_n\, |\, h_{\rm sml})~,
\end{equation}
where $m_n$ is the mass of the $n^{\rm th}$ neighboring particle, $W$ is the kernel weighting function (with units of inverse volume), $N_{\rm ngb}$ is the number of nearest neighbors, $r_n$ is the distance between the point $s$ to the $n^{\rm th}$ particle, and $h_{\rm sml}$ is the smoothing length of the kernel.
The adopted kernel is a top hat, which takes the form
\begin{equation}
    W(r\, |\, h_{\rm sml}) =\begin{cases} 
      \frac3{4\pi h_{\rm sml}^3} &  \, |{r}/{h_{\rm sml}}| \leq 1 \\
      0 & \, {\rm otherwise}
   \end{cases}~.
\end{equation}
The smoothing length is determined by finding the maximum distance to the nearest 32 neighbors using {\sc scipy.spatial.KDTree} \citep{Scipy_2020}.
It should be noted that the top-hat kernel estimates slightly higher densities (by $\sim5\%$) than other common kernel density functions (cubic spline or Gaussian) due to its shape~\citep[see, e.g.,][their Appendix~A]{Qi_2025}; however, this change does not systematically change any of the core results of this work.

We sample radially in $300$ bins ranging from $2.8$ times the softening length~($\sim1.2$ kpc) to $600$ kpc.\ignorespaces
\footnote{\ignorespaces
For distances greater than $\sim2.8\times$ the softening length, gravity is fully Newtonian, whereas below this scale, the shape of the potential is impacted significantly by the choice of softening length.
\edit{
Although previous work \citep[e.g.,][]{Ludlow_2020} suggests that full numerical convergence of dark matter density profiles may require radii somewhat larger than $2.8\times$ the softening, our analysis does not rely on the detailed structure at the smallest resolved radii. 
In practice, many of the dominant differences between hydrodynamical and DMO runs extend to radii closer to $\sim0.02R_{\rm 200}$ ($\sim5$ kpc or $\sim12\times$ the softening length; see, e.g., Figures~\ref{fig:density_profiles},~\ref{fig:comp_to_DMO},~and~\ref{fig:adiabatic_contraction})
}
}
At each radius, we sample this reconstructed density field uniformly along a spherical shell with $10^3$ points.
We take the average density on this shell as the density at that particular radial location.
We also calculate a standard deviation of densities at each radial bin as a measure of azimuthal variations of the density within the shell.
The typical standard deviation of densities is $\lesssim0.1$ dex at all radii within $R_{200}$, suggesting that the halos are reasonably spherical---even while including the contribution from satellites.
We leave a more careful examination of the three-dimensional shape of the dark matter halos for future work.

\subsubsection{Fitting Analytic Profiles}
\label{subsubsec:fitting}

In Section~\ref{subsubsec:shapes}, we fit each profile with an analytic function to get a more quantitative understanding of the shape of the dark matter density profiles.
We fit the profiles with a generalized NFW~(gNFW) profile of the form
\begin{equation}
    \label{eqn:gNFW}
    \rho(r) = \cfrac{\rho_s}{\left(\frac{r}{r_s}\right)^{\gamma}\left[1+\left(\frac{r}{r_s}\right)^{\alpha}\right]^{(\beta-\gamma)/\alpha}}~,
\end{equation}
where $\rho_s$ and $r_s$ are the scale density and radius (respectively), $\gamma$ is the inner slope, $\beta$ is the outer slope, and $\alpha$ regulates how sharply the transition between the two occurs~\citep[e.g.,][]{Jaffe_1983,Hernquist_1990,Zhao_1996,Merritt_2006,DiCintio_2014b}.
In the case that $\alpha=1,\ \beta=3,~{\rm and}~\gamma=1$, the gNFW profile is identical to that of the canonical NFW profile~(i.e., Equation~\ref{eqn:NFW}).

We fit the gNFW profiles using an iterative procedure to obtain priors for the free parameters and ensure convergence.
First, we fit a standard NFW profile to obtain priors on $\rho_s$ and $r_s$ using non-linear least squares minimization ({\sc scipy.optimize.curvefit}; \citeauthor{Scipy_2020} \citeyear{Scipy_2020}).
We then fit a gNFW profile with $\rho_s$ and $r_s$ fixed to the NFW values to derive priors for the shape parameters~($\alpha, \beta, \gamma$; again using non-linear least-squares minimization).
Finally, we perform a full Markov Chain Monte Carlo (MCMC) fit using the {\sc emcee} package~\citep{emcee}, allowing all five parameters to vary simultaneously. 
The MCMC fitting uses a Gaussian likelihood of the form
\begin{equation}
    \log{L}(\phi) = -\frac12 \sum\left(\frac{\log\rho-\log\rho_{\rm model}(r | \phi)}{\sigma_\rho}\right)^2 \, ,
\end{equation}
where $\log\rho$ is the measured average density profile in each radial bin, $\log\rho_{\rm model}(r | \phi)$ is the gNFW prediction at radius $r$ given fit parameters $\phi$, and $\sigma_\rho$ is the standard deviation of densities about the average profile.
We adopt broad Gaussian priors centered on the values obtained from the previous steps using a factor of $5\times$ the uncertainty on the non-linear least squares fits (given by the square root of the covariance matrix along the diagonal) as the width of the Gaussian prior.
This ensures the gNFW fits are not restricted to NFW-like solutions but instead use the NFW fit only to guide initialization.
Finally, we limit this fitting procedure to radii less than $R_{200}$.

\subsection{Emulators}
\label{subsec:emulator}

Throughout this work, we employ a neural network emulator to act as a multi-dimensional interpolator of the parameter space sampled in the DREAMS simulations. These emulators are trained to predict summary statistics of the simulations as a function of the underlying cosmological and astrophysical parameters.
By learning the complex, non-linear mappings between input parameters and simulated outputs, the emulator efficiently explores parameter dependencies and makes predictions without the computational cost of running additional simulations.
The emulator also provides estimates of prediction uncertainty, enabling robust marginalization over multiple parameters at once.
We briefly describe our emulators in this section, noting that they are very similar in both spirit and implementation to that of \citetalias{Rose_2025}.

Our emulators take the five simulation parameters as input ($\bar{e}_w$, $\kappa_w$, $\epsilon_{f,\,\mathrm{high}}$, $\Omega_{\rm M}$, and $\sigma_8$) as well the mass of the halo ($M_{\rm halo}$).
These inputs are fed into a series of fully connected layers, which output a prediction for the mean and standard deviation of the quantity of interest.
The loss function is defined as
\begin{equation}
    \label{eqn:loss_function}
    {\cal L} =\sum_{{\rm batch}}(x-\mu)^2 + \sum_{ {\rm batch}}\left[(x-\mu)^2-\sigma^2\right]^2~,
\end{equation}
where $x$ is the true value of the target in each simulation, $\mu$ is the predicted mean, and $\sigma$ is the predicted standard deviation.
This loss function minimizes the mean-squared-error of the data~(first term) while also ensuring that the measured deviation is reproduced~(second term).
It allows the network to learn both accurate predictions and accurate mean and variance estimation~\citep[][\citetalias{Rose_2025}]{Jeffrey_2020,Paco_2022}.

We split the data into training, validation, and test sets, which comprise $80\%$, $10\%$, and $10\%$ of the dataset, respectively.
We optimize our models using the {\sc optuna} package~\citep{Optuna} to select: (i)~the number of fully connected layers, (ii)~the nodes in each layer, (iii)~the learning rate, (iv)~the weight decay, and (v)~the dropout rate.
Following \citetalias{Rose_2025}, we vary the number of layers from $1$ to $5$, the number of neurons from $4$ to $10^3$, the learning rate from $10^{-5}$ to $10^{-1}$, the weight decay from $10^{-8}$ to $1$, and the dropout rate from $0.2$ to $0.8$.
The hyperparameter tuning is done using $50$ trials consisting of $500$ epochs each.

We employ emulators 
to output gNFW shape parameters (Figures~\ref{fig:gNFW_norm}~and~\ref{fig:gNFW} in Section~\ref{subsec:gNFW}), dark matter mass growth in the inner regions of halos (Figure~\ref{fig:mass_growth} in Section~\ref{subsubsec:mass_growth}), the stellar mass and black hole mass of the central galaxies (Figure~\ref{fig:stellar_mass} in Section~\ref{subsubsec:mass_growth}), the lifetime feedback energy output by AGN (Figure~\ref{fig:appendix_bh_mass} in Appendix~\ref{appendix:AGN_anamoly}), and fitting parameters for our adiabatic contraction calculations (left panel of Figure~\ref{fig:appendix_AC2} in Appendix~\ref{appendix:Aandw}).
In each case, we hyperparameter optimize and train ten instances of the emulator with different random initializations of the weights (for a total of $50$ models created for this work).
In all the results showing predictions from the emulators, we report the mean of the predictions from the ten instances.
Each instance we report in this work is comprised of emulation of $1000$ halos of a single parameter variation keeping the other parameters fixed at their fiducial TNG values and halo mass fixed to $M_{\rm Halo}=10^{12}~{\rm M}_\odot$.
This means that, e.g., for each panel in Figure~\ref{fig:gNFW_norm} there are a total of $10\times1000$ draws from the emulators (for a total of 60,000 in the whole figure).
Figure~\ref{fig:appendix_emulator_validation}  provides an example validation of the emulator that predicts the gNFW shape parameters.

\subsubsection{Definition of Halo-to-Halo Variation}

We quote the total uncertainty on the emulator predictions as
\begin{equation}
    \label{eqn:emulator_uncertainty}
    \sigma_{\rm total}^2 = \bar{\sigma}_{\rm pred}^2 + \sigma_{\rm ensemble}^2 \,,
\end{equation}
where $\bar{\sigma}_{\rm pred}$ is the average predicted uncertainty from each initialization of the emulator, which serves as an estimation of the halo-to-halo variance.
We specifically define the term ``halo-to-halo variance'' as $\sigma^2(X | M_{\rm halo}, \theta)$, where $X$ is the target label (e.g., gNFW shape parameters), $M_{\rm halo}$ is the mass of the halo, and $\theta$ is the vector of simulation features ($\bar{e}_w$, $\kappa_w$, $\epsilon_{f,\,{\rm high}}$, $\Omega_{\rm M}$, and $\sigma_8$).
We thus keep $M_{\rm halo}$ fixed to the fiducial value of $10^{12}~{\rm M_\odot}$ (unless otherwise specified), roughly the median of the DREAMS halo distributions, when making predictions from each emulator so that $\sigma_{\rm pred}^2\equiv\sigma^2(X | M_{\rm halo}, \theta)$. Given the relatively small mass range of this DREAMS suite, the variation in the labels with respect to $M_{\rm halo}$ is usually negligible (although, as we discuss below, not always).
Finally, $\sigma_{\rm ensemble}^2$ of Equation~\ref{eqn:emulator_uncertainty} is the epistemic uncertainty, which captures the uncertainty of the network itself.
In practice, we find that the epistemic uncertainty is always negligible compared to the halo-to-halo variation.
We therefore use the term $\sigma_{\rm total}^2$ interchangeably with the ``halo-to-halo variation''.
Finally, we note that we adopt the convention of quoting halo-to-halo variation as $\pm1\sigma_{\rm total}$ about the mean predictions of the emulators.

\subsection{Halo Weighting Scheme}
\label{subsec:weighting}

\begin{figure*}
    \centering
    \includegraphics[width=0.98\linewidth]{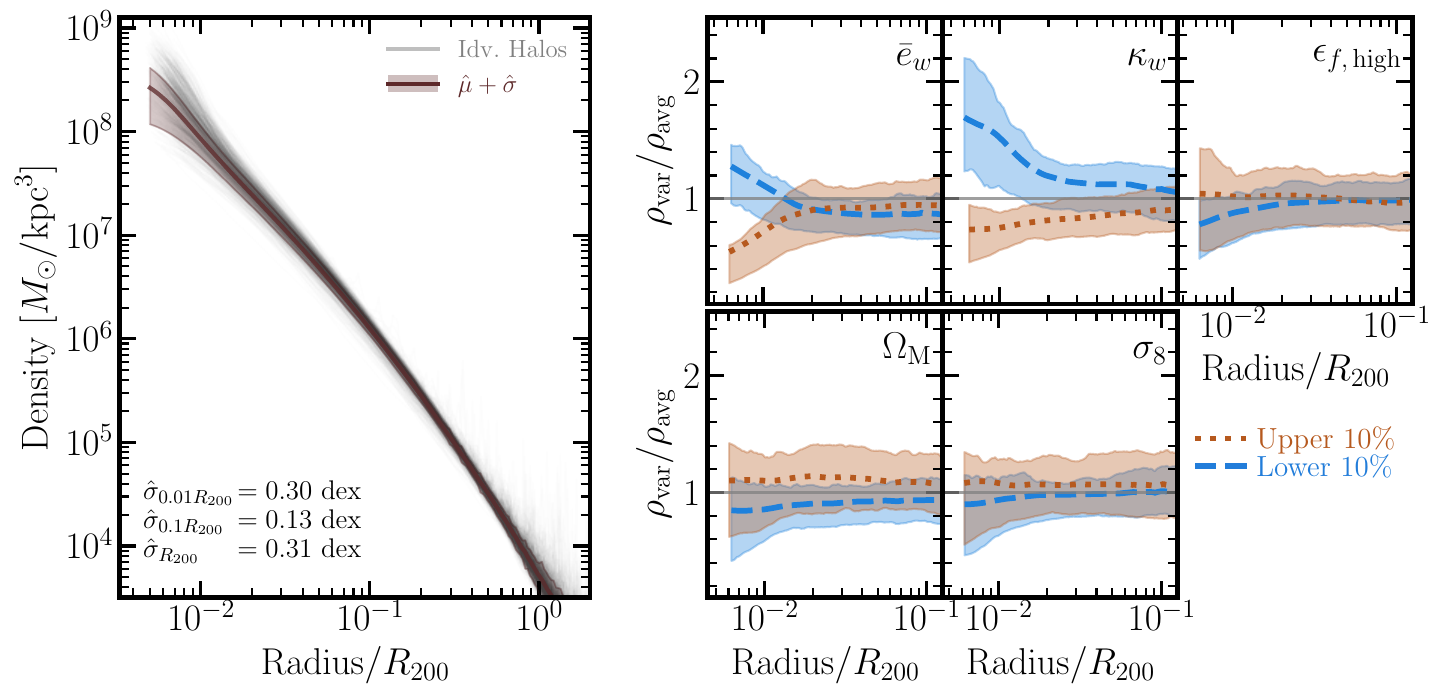}
    \caption{{\bf Dark Matter Density Profiles in the DREAMS CDM Milky Way-Mass Suite.}
    The large left-hand panel shows the dark matter density as a function of radius for all halos in the DREAMS CDM Milky Way-mass suite from $\sim1.2$~kpc to $600$~kpc (normalized by $R_{200}$).
    The \edit{brown} line and shaded region correspond to the weighted mean and plus/minus standard deviation of the individual profiles (see Section~\ref{subsec:weighting} and \protect\citetalias{CDM_Centrals_2025} for details on the weighting procedure).
    The values in the bottom left of the large panel correspond to the weighted standard deviation of densities at $0.01R_{200}$ ($\sim2~{\rm kpc}$), $0.1R_{200}$ ($\sim20$~kpc), and $R_{200}$ ($\sim200$~kpc).
    The panels on the right show how the density depends on the upper~(dotted orange) and lower~(dashed blue) $10\%$ range for each of the parameters (compared to the average $10\%$ of the parameter ranges, $\rho_{\rm avg}$): $\bar{e}_w$~(top left), $\kappa_w$~(top middle), $\epsilon_{f,\,\mathrm{high}}$~(top right), $\Omega_{\rm M}$~(bottom left), and $\sigma_8$~(bottom middle).
    The colored line in each panel is the median density profile, while the colored shaded region represents the $25^{\rm th}$ and $75^{\rm th}$ percentile.
    Overall, the profiles are self-similar, with a modest amount of scatter. 
    Moreover, the baryonic parameters have the strongest effect on the central densities.
    }
    \label{fig:density_profiles}
\end{figure*}

The TNG model has undoubtedly proven to be a widely successful and useful tool for understanding galaxy evolution, yet, as we show in this work, it is not completely robust to the changes in feedback and cosmology of DREAMS.
Put simply, not all of the halos reproduce realistic systems within the feedback variations.
We therefore report all of the main summary statistics~(see left panels of Figures~\ref{fig:density_profiles},~\ref{fig:comp_to_DMO},~and~\ref{fig:adiabatic_contraction}) with weights that reflect whether the halos are consistent with observed galaxies, rather than assuming all halos contribute equally.
All quantities that are weighted will be denoted with the overhead ``hat''~(e.g.,~$\widehat{\sigma}$).

To quantify how reasonable the DREAMS parameter variations are, we follow the halo weighting scheme outlined in \citetalias{CDM_Centrals_2025}.
That work uses a ``pseudo-posterior'' to determine how well a simulated halo's properties compare to well-established scaling relations from observations~(in the case of \citetalias{CDM_Centrals_2025}, the stellar mass--halo mass relation; \citeauthor{SagaV}~\citeyear{SagaV}).
Notably, since the properties of individual halos depend on their detailed formation histories, the weighting is not calculated on a halo-by-halo basis.
Instead, we separate the model parameter space into $30$ evenly-space bins along each dimension ($\bar{e}_w$, $\kappa_w$, and $\epsilon_{f,\,{\rm high}}$)\ignorespaces
\footnote{\ignorespaces
The version of the weighting scheme presented here does not include variations of $\Omega_{\rm M}$ or $\sigma_8$.
Empirically, we find that the weighting scheme does not put strong constraints on these two cosmological parameters, suggesting that they do not play a significant role in setting the stellar mass--halo mass relation.
}
and apply the weights to halos that fall within each bin.
\citetalias{CDM_Centrals_2025} use a normalizing flow emulator \citep[similar to that of][]{Nguyen_2024} to generate samples consistent with the underlying DREAMS data to more finely sample the high-dimensional parameter space.
A sample's residual given some set of input parameters, $\theta$, is calculated such that
\begin{equation}
    R_{\theta_j}\ =\ \frac{1}{N_j}\sum_{n=1}^{N_j} \left(Y_{j,n}^{\mathrm{em}}-g\left(X_{j,n}^{\rm em}\right) \right)~,
\end{equation}
where $N_j$ is the number of samples (set to $10^3$) in each parameter space bin $\theta_j$, $X_{j,n}^{\rm em}$ and $Y_{j,n}^{\rm em}$ are the emulated quantities (halo mass and stellar mass, respectively) for the $n^{\rm th}$ sample, and $g(X)$ is a piece-wise linear parameterization of the \cite{SagaV} stellar mass--halo mass relation.
When a halo (or emulated sample) is near the observed stellar mass--halo mass relation, its $R_{\theta_j}$ will be close to $0$.
The weight associated with parameter bin $\theta_j$ is then calculated as
\begin{equation}
    \tilde{w}_j = \mathrm{exp} \left( - \frac{R^2_{\theta_j}}{2 \tau ^2} \right)~,
\end{equation}
where $\tau$ is a free parameter 
set to $0.2~\log[{\rm M}_\odot]$ (see \citetalias{CDM_Centrals_2025}).
We note that the results of this work do not depend sensitively on the choice of $\tau$.
Finally, we normalize the weights such that 
\begin{equation}
    w_j=\frac{\widetilde{w}_j}{\sum_j \widetilde{w}_j } \, .
\end{equation}
\edit{The normalized weights admit a natural interpretation as importance weights \citep{farahi2026simulation}.}  
The results of this weighting procedure can be seen in Figure 1 of \citetalias{CDM_Centrals_2025}.
To briefly summarize, we down-weight the extreme values of $\bar{e}_w$ and $\kappa_w$ and the lower-than-fiducial values of $\epsilon_{f,\,\mathrm{high}}$, but do not place strong constraints on $\Omega_{\rm M}$ or $\sigma_8$. 

We note that our main conclusions are not particularly sensitive to using this weighting scheme since, as we will show, the density profiles themselves are not strongly dependent on the supernova and AGN feedback parameters~(see discussion in Section~\ref{subsec:density_profiles}).
However, since there are halos with unrealistically large/small stellar masses given the halo mass of the Milky Way, we aim to down weight their contributions.

\section{Results}\label{sec:results}

\subsection{Dark Matter Density Profiles}
\label{subsec:density_profiles}

Figure~\ref{fig:density_profiles} shows the dark matter density profiles for all of the DREAMS halos~(large panel on left).
To compare the results across simulations, the radii are normalized by $R_{200}$ for each galaxy.
The weighted mean and standard deviation are shown by the \edit{brown} line and shaded band, respectively; these correspond to the density profiles of those Milky Way halos whose central galaxies are most consistent with the observed stellar mass--halo mass relation.

On the whole, the dark matter profiles are largely consistent in terms of normalization, which is, in part, by construction of the selection function of the DREAMS halos.
There is a tight (weighted) spread in densities of $0.13$~dex at $0.1R_{200}$~($\sim20~{\rm kpc}$).
At larger radii, the spread increases by a factor of \edit{$\sim1.5$~(increasing from $0.13~{\rm dex}$ to $0.31~{\rm dex}$ at $R_{200}$)} which is a feature of the halos containing many subhalos at these large radii.
Note that the presence of more satellites leads to occasional ``spikes'' at large radii.
In the inner-most regions of the halo, $0.01R_{200}$~($\sim 2~{\rm kpc}$), there is a much wider distribution of densities with a spread of $0.31~{\rm dex}$~(a factor of $\gtrsim2$ increase from $0.1R_{200}$). 
Interestingly, this trend is consistent with massive halos with mass above $10^{13}~{\rm M}_\odot$ \citep[see left panel of Figure 2 in][]{Farahi_2022profile}. 

The variation in the inner regions of the halo is primarily driven by the supernova physics.\ignorespaces
\footnote{\ignorespaces
Since these are the raw simulation outputs, deviations could also be driven by the combination of the supernovae parameters and halo-to-halo variance.
To that end, we probe the one-parameter dependencies more directly in the following section using our neural network emulator.
}
To show this concretely, each panel on the right of Figure~\ref{fig:density_profiles} shows the profiles broken down into the upper~(dotted orange) and lower~(dashed blue) $10\%$ of each parameter distribution, normalized average density of the middle $10\%$ of the parameter ($\rho_{\rm avg}$).
These percentiles represent the $\sim100$ halos with the highest~(or lowest) values of the parameter, while marginalizing over the others.
For each band, the line represents the median profile while the spread indicates the inner-quartile range~($25^{\rm th}$ and $75^{\rm th}$ percentiles).
The top row shows the astrophysical parameter variations: $\bar{e}_w$ in top left, $\kappa_w$ in top middle, and $\epsilon_{f,\,\mathrm{high}}$ in the top right. 
The bottom row shows the cosmological parameter variations: $\Omega_{\rm M}$ in the bottom left and $\sigma_8$ in the bottom middle. 


\edit{At the inner most radii ($0.06R_{200}$), t}he lower $\bar{e}_w$ values correspond to a median density ratio of $1.278_{-0.315}^{+0.185}$ ($\sim30\%$ increase), while the upper $\bar{e}_w$ values have $0.737_{-0.281}^{0.213}$ ($\sim30\%$ decrease).
However, within the spread, the lower $\bar{e}_w$ variations are consistent with no significant change in the inner density.
For $\kappa_w$, the lower values correspond to a ratio of $1.697^{+0.508}_{-0.463}$ ($\sim70\%$ increase), and the upper values correspond to a ratio of $0.737^{+0.213}_{-0.281}$ ($\sim30\%$ decrease).
Notably, neither the upper nor lower $\kappa_w$ ranges are consistent with the average profile, at least within the inner-quartile range.
Interestingly, the enhancement/decrease caused by $\kappa_w$ (which controls wind speed in our simulations) extends to much larger radii than those of $\bar{e}_w$ (which controls wind energy).

In terms of $\epsilon_{f,\,\mathrm{high}}$, the upper values are consistent with the average profile (ratio of $1.045_{-0.486}^{+0.387}$), while the lower values are very slightly below the average profile (ratio of $0.782_{-0.295}^{0.207}$).
While very slight, this decrease in central density with decreased $\epsilon_{f,\,\mathrm{high}}$ is likely related to a self-regulation of the black holes which we discuss in more detail in Section~\ref{subsubsec:mass_growth}.

For both the $\Omega_{\rm M}$ and $\sigma_8$ variations, both the lower and upper ranges are consistent with the average profile \edit{within the uncertainty on the median.}
\edit{It is worth appreciating that if we were to consider wider variations on $\Omega_{\rm M}$ and $\sigma_8$ as in the CAMELS simulations~(\citeauthor{CAMELS_2021}~\citeyear{CAMELS_2021}; $\Omega_{\rm M}\in[0.1,0.5]$ and $\sigma_8\in[0.6,1.0]$)
it is possible that deviations from the average profile might be more significant.
}

\subsection{Analytic Fits to Profiles}
\label{subsec:gNFW}

\begin{figure*}
    \centering
    \includegraphics[width=0.98\linewidth]{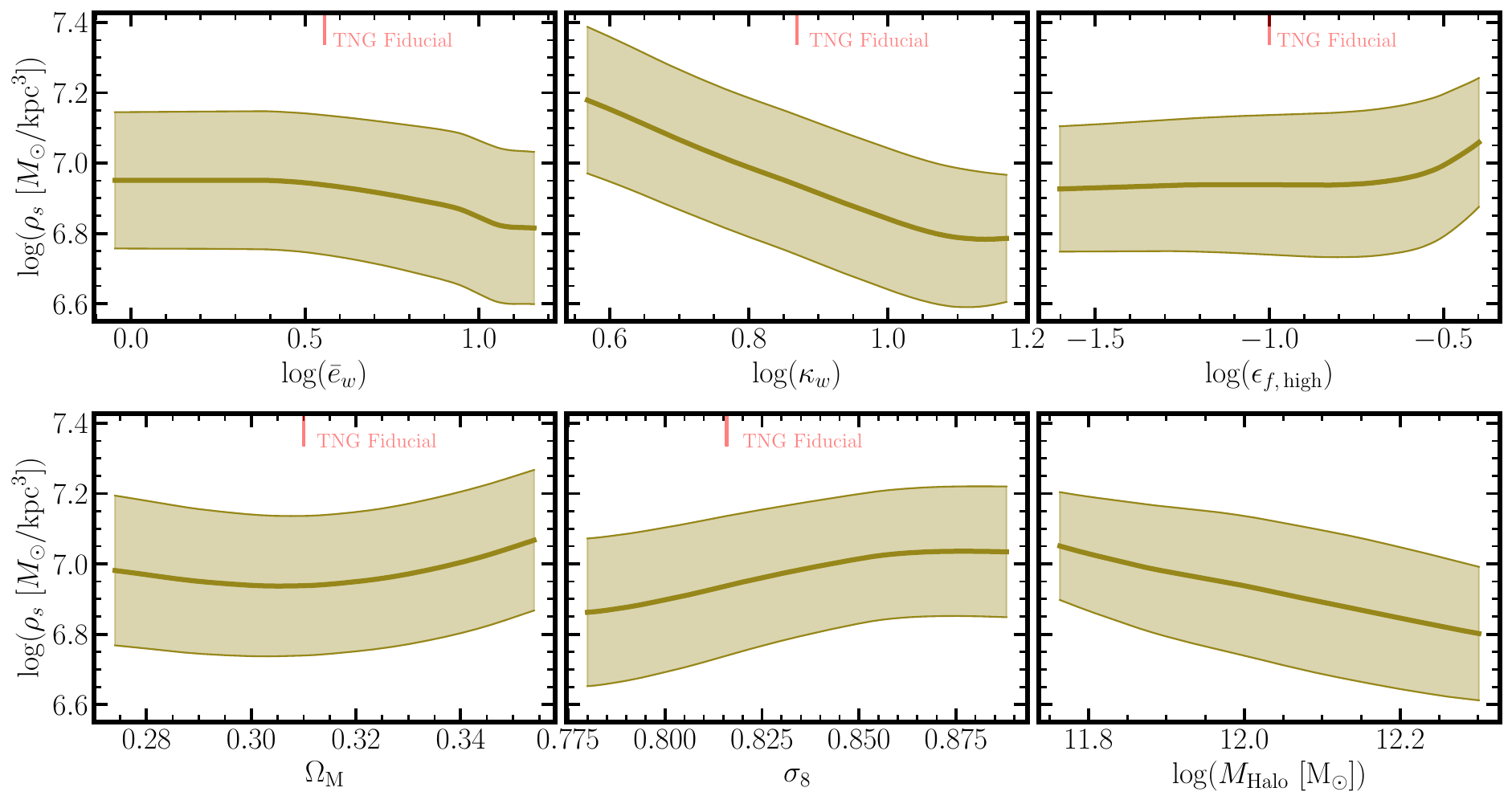}
    \caption{{\bf Dependence of generalized NFW scale density ($\rho_s$) on Astrophysics, Cosmology, and Halo Mass.}
    Predictions from an ensemble of emulators (see Section~\ref{subsec:emulator}) for the scale density, $\rho_s$, of the DREAMS CDM Milky Way-mass density profiles.
    We show the predictions made by the emulator as a function of $\bar{e}_w$~(top left), $\kappa_w$~(top middle), $\epsilon_{f,\,\mathrm{high}}$~(top right), $\Omega_{\rm M}$~(bottom left), $\sigma_8$~(bottom middle), and $M_{\rm halo}$ (bottom right).
    The shaded regions represent the one standard deviation uncertainty of the predictions based on the individual model prediction uncertainty as well as variance across the emulators~(via Equation~\ref{eqn:emulator_uncertainty}).
    The vertical red solid line at the top of each panel corresponds to the fiducial TNG value~(see Table~\ref{tab:dreams_parameters}).
    In every panel, the non-varied parameters are fixed to their fiducial values and $M_{\rm Halo} = 10^{12}~{\rm M}_\odot$ unless explicitly varied.
    While not shown, the emulator also predicts the scale radius.
    The trends for the scale radius are identical to those shown here, just in the opposite direction (i.e., increasing $\kappa_w$ increases $r_s$).
    We find that only $\kappa_w$ drives variation in $\rho_s$ more significant than the halo-to-halo variation.
    }
    \label{fig:gNFW_norm}
\end{figure*}
\begin{figure}
    \centering
    \includegraphics[width=\linewidth]{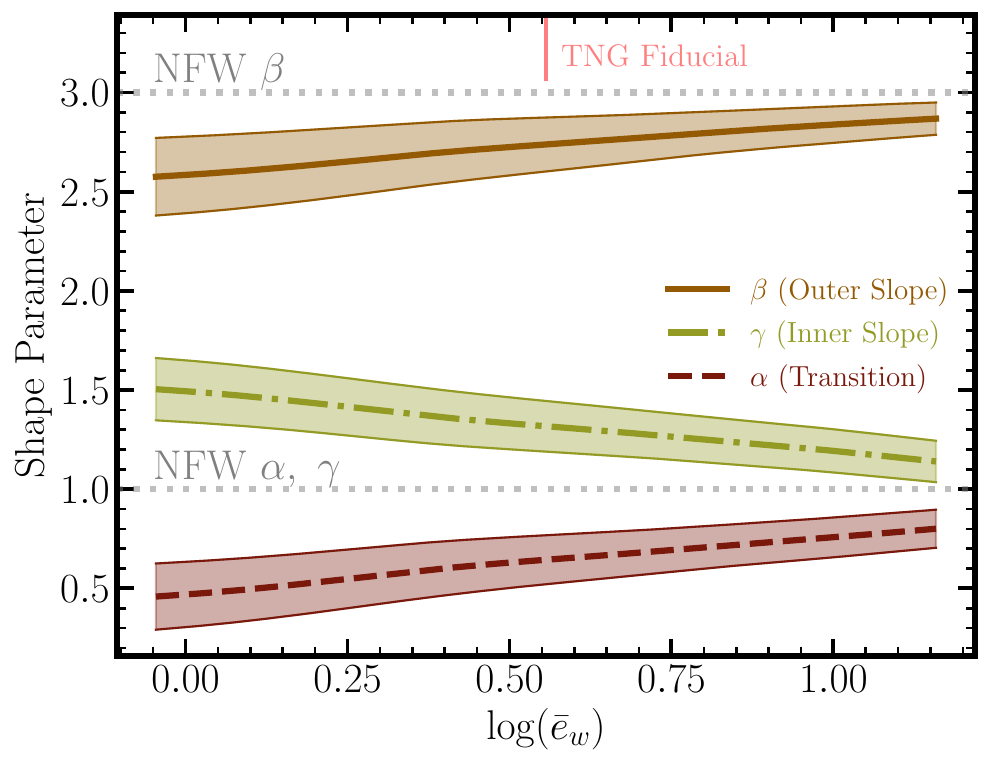}
    \caption{
    {\bf Dependence of generalized NFW shape parameters on $\mathbf{\bar{e}_w}$.}
    Predictions from our neural network emulator for the dependence of the inner slope ($\gamma$; dot-dashed line), transition rate ($\alpha$; dashed line), and outer slope ($\beta$) of the best-fit gNFW profiles on the supernova wind energy ($\bar{e}_w$).
    As a point of reference, the dotted gray lines show the canonical NFW profile values ($\alpha=\gamma=1$ and $\beta=3$).
    The short vertical solid line at the top corresponds to the fiducial TNG value of $\bar{e}_w=3.6$~(see Table~\ref{tab:dreams_parameters}; Section~\ref{subsec:astro_variations}).
    The average prediction lines are the mean of our ten emulators while the shaded regions approximate the halo-to-halo variation (although there is also a small contribution from the combination of emulators, see Section~\ref{subsec:emulator} for details).
    We note that, within the uncertainty due to halo-to-halo variation, the shape parameters have no strong dependence on any of the other DREAMS simulation parameters or halo mass.
    }
    \label{fig:gNFW}
\end{figure}

\subsubsection{Scale Density}
\label{subsubsec:rho_0}

We now examine how the best-fit gNFW parameters change as a function of the DREAMS astrophysics and cosmology variations.
Figure~\ref{fig:gNFW_norm} shows the scale density $\rho_s$ as a function of the five parameters, as well as the halo mass.
More specifically, we show the predictions made by an emulator trained to learn the one-parameter dependence of $\rho_s$~(see Section~\ref{subsec:emulator} for details on the training and hyperparameter optimization).
The advantage of this approach, compared to showing all the fits to the profiles in Figure~\ref{fig:density_profiles}, is that we can directly parameterize the dependence of $\rho_s$ on a parameter-by-parameter basis, instead of sampling from realizations that have multi-parameter dependencies.
We therefore keep the parameters fixed to their fiducial TNG value and halo mass fixed to $10^{12}~{\rm M}_\odot$, except when explicitly varied.
The solid lines are the average predictions of the ensemble of ten emulators.
Each shaded region represents the $1\sigma$ uncertainty of the ensemble of emulators via Equation~\ref{eqn:emulator_uncertainty}, which gives a proxy for the intrinsic halo-to-halo scatter at a given parameter point.
Finally, the vertical solid red line at the top of each panel marks the fiducial TNG parameter~(from Table~\ref{tab:dreams_parameters}) and the default halo mass.

We find that $\rho_s$ has some dependency on all five of the simulation parameters and halo mass, albeit  to a different extent in each case.
The parameter that holds the most importance in setting the scale density is $\kappa_w$~(top-middle panel).
In fact, $\log\rho_s$ is roughly inversely proportional to $\log \kappa_w$, ranging from values of $\rho_s=10^{7.2}~{\rm M}_\odot\,{\rm kpc}^{-3}$ at the lowest $\kappa_w$ values to $\rho_s=10^{6.8}~{\rm M}_\odot\,{\rm kpc}^{-3}$ at the highest values of $\kappa_w$, a change of $\sim0.4$ dex.
In comparison, the average halo-to-halo variance for $\kappa_w$ is also $\sim0.4$ dex (shaded region).
Thus, the impact of $\kappa_w$ is comparable to that of the intrinsic scatter of the sample.
Beyond $\kappa_w$, none of the other simulation parameters drive variations comparable to the average halo-to-halo variations.
The total variation in the mean relation for $\bar{e}_w$ is $\sim0.1$~dex, $\epsilon_{f,\,\mathrm{high}}$ is $\sim0.2$~dex (most of which is at the high AGN feedback variations, see discussion in Section~\ref{subsubsec:mass_growth}), $\Omega_{\rm M}$ is $\sim0.1$~dex, and $\sigma_8$ is $\sim0.15$~dex.
Compared to the typical halo-to-halo variation of $\sim0.2$ dex, we conclude that none of these parameters is crucial in setting the normalization of the dark matter density profiles.
Similarly, despite there being a somewhat coherent inverse relationship between halo mass and $\log(\rho_s)$~(bottom-right panel of Figure~\ref{fig:gNFW_norm}), halo mass does not drive variations more significant than halo-to-halo variation over the range of our Milky Way-mass halos.


In summary, the halo-to-halo variation is the dominant driver of the normalization of the dark matter density profiles for Milky Way-mass systems, at least within the TNG model.
With the exception of supernova wind speed ($\kappa_w$), none of the DREAMS simulation parameters, nor halo mass, drive variations comparable to the halo-to-halo variations.
Finally, while not shown, we obtain qualitatively similar trends for the scale radius $r_s$ (although in the opposite direction as $r_s$ and $\rho_s$ are inversely proportional to each other).

\subsubsection{Profile Shape Parameters}
\label{subsubsec:shapes}

Figure~\ref{fig:gNFW} shows the best-fit shape parameters $\beta$~(outer slope), $\gamma$~(inner slope) and $\alpha$~(transition rate) for the gNFW profile as a function of $\bar{e}_w$. 
Within the uncertainty of halo-to-halo scatter, the shape parameters do not depend sensitively on any other parameter, so they are not shown in this figure.
The dotted gray lines on Figure~\ref{fig:gNFW} represent $\alpha=\gamma=1$ and $\beta=3$, consistent with a canonical NFW profile.

When increasing $\bar{e}_w$, all the shape parameters tend towards the canonical NFW values.
Recall that the NFW profile is the empirical prediction from DMO simulations~\citep*{NFW_1997}.
What is happening here is that these extremely high wind energy variations of the TNG model prevent the growth of any significant stellar component (which we show explicitly in Section~\ref{subsubsec:mass_growth}), thus the halos are less dominated by the presence of baryons and behave more like collisionless DMO simulations.
It should be noted, though, that the gNFW shape parameters only tend towards the NFW values and do not actually reach them.
Thus, while the stronger feedback is preventing the growth of the stellar mass, it is not entirely prevented.
We explore the stellar mass growth (or lack thereof) further in the next section and in \cite{CDM_Satellites_2025}.

It is interesting to note that the variations to the supernova wind {energy}~($\bar{e}_w$) seem to drive the shape of the dark matter density profiles, while changes to the supernova wind {speed}~($\kappa_w$) have a stronger impact on the normalization of the profile.
This is likely related to the effects of wind speeds extending to larger radii, where $\rho_s$ is determined (see central panel on right-hand side of Figure~\ref{fig:density_profiles}).
It is also possible that ejecting winds faster~(as is the case with $\kappa_w$) can more efficiently disrupt the potential, changing the dark matter density, compared to increases to the total feedback energy~($\bar{e}_w$) which is better at suppressing the growth entirely of the baryon component of the galaxy.

Overall, the single most important feature in setting the normalization and shape of density profiles for Milky Way-mass halos in the TNG model is halo-to-halo variance.
Beyond halo-to-halo scatter, extreme variations to supernova feedback can slightly change the overall normalization~($\kappa_w$) or its shape~($\bar{e}_w$) beyond that of the intrinsic scatter.

\subsection{Central Dark Matter Mass Growth}
\label{subsubsec:mass_growth}

\begin{figure*}
    \centering
    \includegraphics[clip,trim={0 8.41cm 0 0},width=0.98\linewidth]{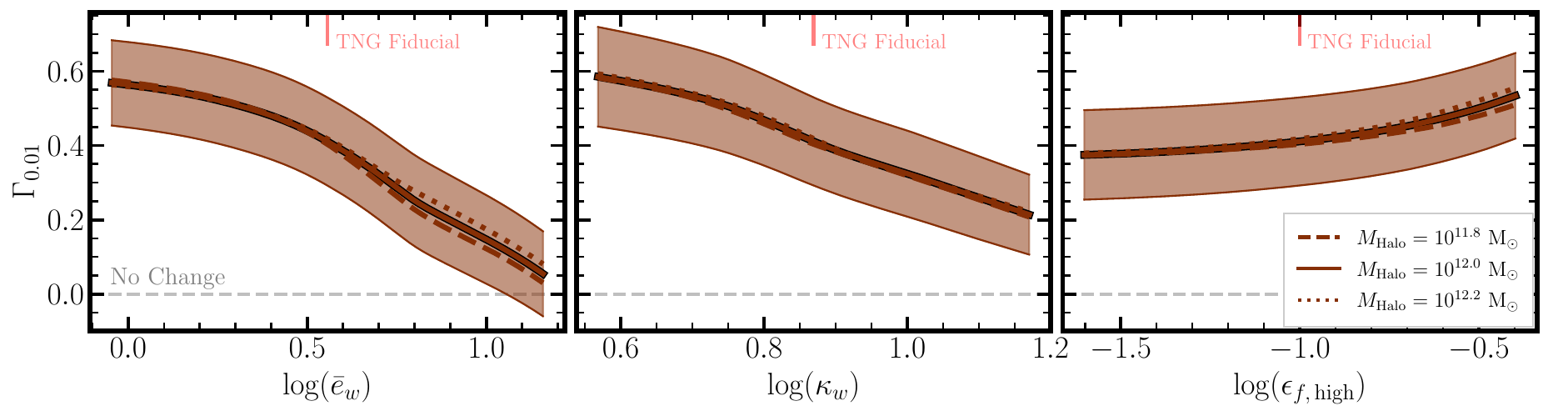}
    \caption{{\bf Central Mass Growth ($\mathbf\Gamma$) of Halos at $\mathbf{0.01R_{200}}$.}
    Predictions from our emulators for the central mass growth, defined as the ratio of the dark matter mass in the hydro simulations to that of the DMO simulations~(see Equation~\ref{eqn:mass_growth}) at a radius of $0.01 R_{200}$, as a function of the three DREAMS baryon feedback parameters.
    The solid line represents the average prediction of the parameter from the ensemble of emulators for $M_{\rm Halo} = 10^{12}~{\rm M}_\odot$, while the shaded region is the uncertainty on the predictions as a proxy for halo-to-halo variance~(via Equation~\ref{eqn:emulator_uncertainty}).
    The dashed gray line represents a mass ratio of unity, where the halo mass is unchanged from the hydro to the DMO simulation.
    The short solid lines at the top of each panel correspond to the fiducial TNG value~(see Table~\ref{tab:dreams_parameters}).
    We find that the supernova feedback parameters decrease the central mass growth, whereas the AGN feedback parameter increases the central mass growth.   The dashed and dotted lines show the average for halo masses of $10^{11.8}$ and $10^{12.2}~{\rm M}_\odot$, respectively (corresponding bands not shown).  
    }
    \label{fig:mass_growth}
\end{figure*}

The previous subsection quantified the relative role that baryon feedback, cosmology, and intrinsic halo-to-halo variation play on the overall shape and normalization of the DREAMS dark matter halos.
Here, we further assess the relative contribution of these three features by using the ``central mass growth'' of the halos, 
\begin{equation}
    \label{eqn:mass_growth}
    \Gamma_{0.01} = \log\left[\frac{M_{\rm enc,\,Hydro}(0.01R_{200})}{M_{\rm enc,\,DMO}(0.01R_{200})}\right]~,
\end{equation}
defined as the logarithmic ratio of dark matter mass enclosed at $0.01R_{200}$ from hydro-to-DMO~(similar to the metric used in \citeauthor{Rose_2023} \citeyear{Rose_2023}).
This quantity provides a metric for the changes in the inner dark matter distribution relative to a DMO baseline.
Dark matter will contract because of the presence of the baryons (see also discussion in Section~\ref{subsec:adiabatic_contraction}), and thus $\Gamma_{0.01}$ should be sensitive to the detailed behavior set by the feedback variations.

Figure~\ref{fig:mass_growth} shows the central mass growth as a function of the three DREAMS baryonic feedback parameters.
As in Section~\ref{subsec:gNFW}, we train an ensemble of emulators to learn the single parameter dependency of the central mass growth~(see Section~\ref{subsec:emulator} for more details on the emulator training process) and report the average (solid line) and halo-to-halo variation (shaded region) for $M_{\rm Halo} = 10^{12}~{\rm M}_\odot$. 
All three astrophysics parameters impact the central mass growth of the halos.
The total halo-to-halo variation is $\sim0.2$ dex~(i.e., $\pm0.1$ dex standard deviation), whereas we find variations of $\sim0.5$ dex, $\sim0.4$ dex, and $\sim0.2$ dex for $\bar{e}_w$, $\kappa_w$, and $\epsilon_{f,\,\mathrm{high}}$, respectively.
Conversely, there are no strong trends with either $\Omega_{\rm M}$ or $\sigma_8$ ($\lesssim0.05$ dex) and thus do not report them in Figure~\ref{fig:mass_growth}.
Similarly, variations to the halo mass have negligible impact on the results; Figure~\ref{fig:mass_growth} also shows the average trend for $M_{\rm Halo}=10^{11.8}~{\rm M}_\odot$ (dashed line) and $10^{12.2}~{\rm M}_\odot$ (dotted line), noting that these mass ranges have comparable halo-to-halo variance as the $10^{12.0}~{\rm M}_\odot$ range.

The lowest $\bar{e}_w$ values cause a central mass increase of $\sim0.55$~dex (a factor of $>3$) while the highest $\bar{e}_w$ values only increase the central mass by $\sim0.05$~dex.
The decrease in mass growth from the lowest-to-highest $\bar{e}_w$ values is not linear, however.
Around the fiducial TNG value~($\log \bar{e}_w = 0.55$), the central mass growth decreases more rapidly with increasing $\log \bar{e}_w$ until $\log \bar{e}_w\sim0.7$.
The $\kappa_w$ variations show a similar trend to $\bar{e}_w$, albeit with a less sharp decrease at high values.

\begin{figure*}
    \centering
    \includegraphics[width=0.98\linewidth]{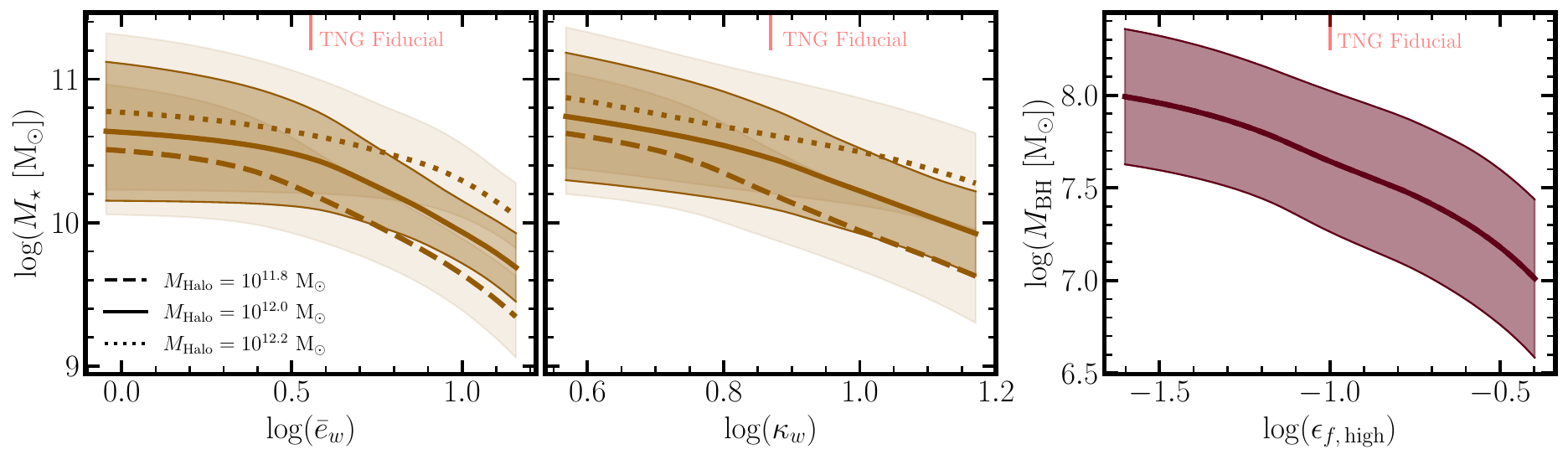}
    \caption{{\bf Properties of Central Galaxy.}
    Predictions from our emulator for the scaling of the stellar mass (left two panels) and black hole mass (right panel) of the Milky Way-mass halo's central galaxy with the DREAMS astrophysics variations.
    The three lines represent predictions from our ensemble of emulators at $M_{\rm Halo}=10^{11.8}~{\rm M}_\odot$~(dashed), $10^{12.0}~{\rm M}_\odot$~(solid), and $10^{12.2}~{\rm M}_\odot$~(dotted).
    The shaded regions represent the halo-to-halo variation on each quantity~(which we find to be similar at each halo mass).
    Increased supernova feedback (both wind energy and speed) decreases the stellar mass of the central galaxy, while increased AGN feedback actually {\it decreases} the central supermassive black hole.
    }
    \label{fig:stellar_mass}
\end{figure*}

Both the $\bar{e}_w$ and $\kappa_w$ trends can be explained by looking at how they affect the stellar mass of the central galaxy---see the left two panels of Figure~\ref{fig:stellar_mass}.
In the low $\bar{e}_w$ and $\kappa_w$ regime, the feedback may no longer be strong enough to prevent gas from further collapsing, thus all the baryons that would have turned into stars eventually do.
Interestingly, it appears that the saturation limit for feedback regulating further star formation is just less than the fiducial TNG values of $\bar{e}_w$ and $\kappa_w$, potentially owing to the fixed star formation efficiency in the TNG model.
At the other end, the strong supernova feedback likely disrupts gas that would have otherwise formed stars and removes it from the system entirely.
The net effect of these disruptions over the lifetime of the galaxy leads to a decrease in its overall stellar mass.
Regardless of the exact physical mechanism, it is clear that the conclusion from previous sections holds here as well: stronger feedback~($\bar{e}_w$ in particular) prevents baryon formation sufficiently easily that the simulations are nearly DMO.

The trends in mass growth with supernova physics are qualitatively similar in different halo mass hosts, with normalization shifts associated with evolution in the stellar mass--halo mass relation (shown in Figure~\ref{fig:stellar_mass} as a dashed line for $M_{\rm Halo}=10^{11.8}~{\rm M}_\odot$, solid for $10^{12.0}~{\rm M}_\odot$, and dotted for $10^{12.2}~{\rm M}_\odot$; see also \citetalias{CDM_Centrals_2025}).
In more detail, the variation in stellar mass is slightly larger for less massive hosts: e.g., $>1$ dex in $10^{11.8}~{\rm M}_\odot$ halos and $\sim0.8$ dex in $10^{12.2}~{\rm M}_\odot$ halos for $\bar{e}_w$ variations, with similar trends, albeit to a lesser extent, with $\kappa_w$.
Thus, both the energy and wind speed variations are less effective at reducing the stellar mass of the central galaxy in more massive hosts, likely because ejecting material out of a deeper potential via supernova feedback is more difficult.
Regardless, the difference in stellar mass growth in the low- versus high-mass halos for supernova energy variations ($\bar{e}_w$) likely also explains the (very minor) differences seen in the $\Gamma_{0.01}$ ratios at high $\bar{e}_w$ arising from variations in halo mass.

The AGN variations follow a different trend, on the other hand.
With increasing $\log \epsilon_{f,\,\mathrm{high}}$, there is a roughly linear increase in the central mass growth; ranging from $\sim0.35$~dex to $\sim0.5$~dex (with no halo mass dependence).
One interpretation is that the increase in AGN feedback may be having a self-regulatory effect~\citep{Ni_2023}.
Increasing AGN feedback strength more efficiently removes gas from the central regions, limiting the gas left to accrete, and driving down the total mass of the black hole over its lifetime (right panel of Figure~\ref{fig:stellar_mass}).
Integrating Equation~\ref{eqn:agn} to obtain the total thermal energy that the AGN outputs over its lifetime, we find the energy output scales with $\epsilon_{f,\,\mathrm{high}}$, but also the total~(accreted) mass of the black hole such that
\begin{equation}
    \label{eqn:agn_lifetime_feedback}
    E_{\rm AGN,\,lifetime} = \epsilon_r \epsilon_{f,\,\mathrm{high}}M_{\rm BH} c^2~.
\end{equation}
Thus, higher values of $\epsilon_{f,\,\mathrm{high}}$ feedback would not output more total thermal energy over their lifetimes than lower values~(shown explicitly in Appendix~\ref{appendix:AGN_anamoly}, Figure~\ref{fig:appendix_bh_mass}).
This limiting efficacy of AGN would also be consistent with the (slight) density enhancement seen in the inner regions in the top-right panel of Figure~\ref{fig:density_profiles}.

\section{Discussion}
\label{sec:discussion}

\subsection{Dark Matter-Only Simulations}
\label{subsec:dmo}

\begin{figure*}
    \centering
    \includegraphics[width=0.98\linewidth]{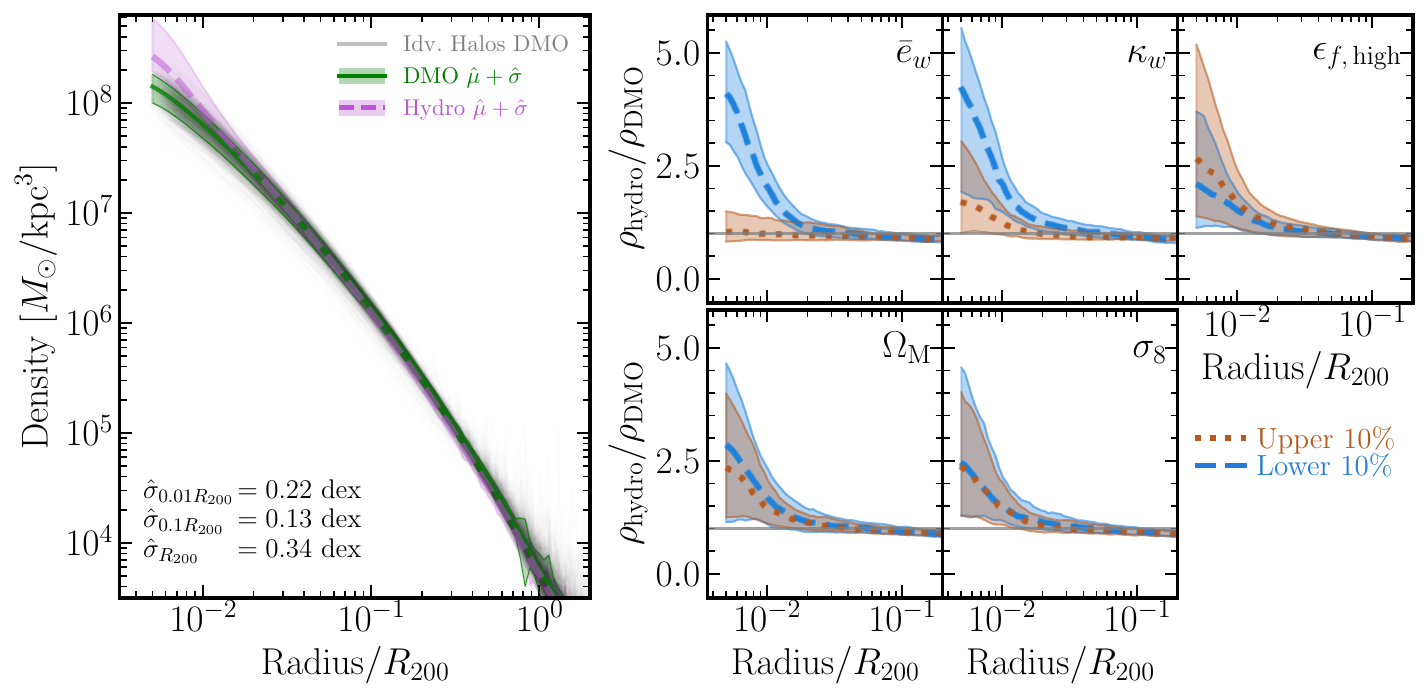}
    \caption{{\bf Density Profiles from Dark Matter-Only Simulations.}
    The left-hand panel is the same as Figure~\ref{fig:density_profiles}, but for the Dark Matter-Only~(DMO) simulations. It shows the DMO density profiles for each simulation normalized by $R_{200}$.
    The green solid line and shaded region is the weighted mean and standard deviation (see Section~\ref{subsec:weighting}) from the DMO simulations, while the dashed pink line and shaded region is the weighted mean and standard deviation from the hydro simulations (i.e., the same as in Figure~\ref{fig:density_profiles}).
    The right-hand panels all show the ratio of the hydrodynamic simulations' density to the corresponding DMO simulation for the same parameter ranges as the right-hand panels of Figure~\ref{fig:density_profiles}.
    We find that supernova feedback variations drive the deviations (or lack thereof) from hydrodynamic to DMO simulations.
    }
    \label{fig:comp_to_DMO}
\end{figure*}

As mentioned in Section~\ref{subsec:DREAMS}, each  hydrodynamic simulation in DREAMS has a corresponding DMO counterpart, which has the equivalent initial conditions.
We can therefore make systematic comparisons on the impact of baryons as a function of the varied parameters in the simulations.

The left-hand panel of Figure~\ref{fig:comp_to_DMO} shows the density profiles from the DMO DREAMS simulations.
Each thin gray line is an individual halo, while the solid green line and shaded region are the mean and standard deviation of the individual profiles. 
For comparison purposes, each DMO halo is assigned the same weight as its hydro counterpart, despite the baryonic feedback parameters not impacting the DMO simulations. 
As a point of reference, we also overplot the densities for the hydro simulations rom Figure~\ref{fig:density_profiles}.
Both the mean and spread of the DMO density profiles are generally less than that of the hydro simulations in the inner radii but are roughly consistent at $\gtrsim0.1R_{200}$.
The spread at $0.01R_{200}$ is $0.22$~dex, which decreases to $0.13$~dex at $0.1R_{200}$ before increasing back to $0.34$~dex at $R_{200}$ due to the presence of subhalos.
The larger spread at small radii cements the notion that the feedback variations are the dominant drivers of density profile variations in these inner regions.
Moreover, the agreement at larger radii ($>0.1R_{200}$) further suggests that intrinsic halo-to-halo variations drive the scatter in this regime.
\ignorespaces
\footnote{\!~\edit{We note that we normalize the radial profiles of both the DMO and hydrodynamic simulations by $R_{200}$.
In detail, $R_{200}$ is not the same radius in the pairs of simulations; however, we find that our results are qualitatively unimpacted by this difference.
We find similar results when normalizing both profiles by their respective best-fit $\rho_0$ and $r_s$ gNFW parameters.
}}

Each of the right-hand panels of Figure~\ref{fig:comp_to_DMO} shows the ratio of the hydro simulation density to the corresponding DMO simulation for different parameters~(top left-to-right: $\bar{e}_w$, $\kappa_w$, and $\epsilon_{f,\,\mathrm{high}}$, bottom left-to-right: $\Omega_{\rm M}$ and $\sigma_8$).
Similar to Figure~\ref{fig:density_profiles}, we show only the median predictions from the upper~(dotted line) and lower~(dashed line) $10\%$ of each of the parameters for clarity.
The shaded regions represent the $25^{\rm th}$ and $75^{\rm th}$ percentiles of the distributions.

In general, the density in the central region of the halo is significantly smaller for the DMO simulations relative to their hydro counterparts.
The ratio of $\rho_{\rm hydro}$ to $\rho_{\rm DMO}$ at small radii ($\lesssim 0.02R_{200}$) indicates a steeper inner profile in the hydro simulations---in agreement with the results of Section~\ref{subsec:gNFW} and from the EAGLE simulations~\citep{Schaller_2015b}.
At larger radii ($\gtrsim0.1R_{200}$), the hydro and DMO profiles appear roughly consistent.
The degree to which the density increases in the inner region is sensitive to the supernova feedback parameters.
In particular, the supernova wind energy~($\bar{e}_w$) has a very strong impact relative to the DMO simulations.
The lower range of $\bar{e}_w$ values leads to a factor of $\sim3$--$4$ increase in the density compared to DMO, whereas the results for the upper range are roughly consistent with the DMO simulations.
This makes sense because the high-energy supernova should prevent baryon content from forming, leading to similar results as the DMO simulations.
The supernova wind speed~($\kappa_w$) shows a similar trend, albeit to a smaller extent, of increased feedback driving the density profiles closer to the DMO simulations.
The upper $10\%$ of $\log(\kappa_w)$ increases the central density by only a factor of $\sim1.5$ on average, whereas the highest $10\%$ change the density by a factor of $>4$.
On the other hand, the AGN feedback and cosmology variations do not have any significant impact relative to the DMO simulations.

We note that the ratio of $\rho_{\rm hydro}/\rho_{\rm DMO}$ only rarely dips below unity in the inner radii of any of our halos, which contrasts results from the FIRE-2 simulations by~\cite{McKeown_2022}, which show that the addition of baryon physics can suppress the inner density by factors of $\sim2$.
The difference between our work and~\cite{McKeown_2022} likely stems from the implementation of feedback in FIRE-2.
One key difference in the FIRE model is the existence of strong, time-variable stellar feedback-driven winds~\citep{Muratov_2015}, which can perturb the galactic potential and ``core'' density profiles~\citep{Governato_2010,Pontzen_2012,Mostow_2024}.
We discuss the role of bursty feedback more in Section~\ref{subsec:bursty}.

\subsection{Contraction of Dark Matter Halos}
\label{subsec:adiabatic_contraction}

As mentioned in the Introduction, adiabatic contraction describes the dark matter's response to baryons cooling within a halo. In short, as the baryons condense in the center of a galaxy, they also pull the dark matter inward, increasing the density in the central region.
Since the DREAMS suite contains both DMO and hydrodynamic realizations of the same halo, it is natural to quantify whether the differences between the two are consistent with an adiabatically contracting halo.

\cite{Blumenthal_1986} suggests that adiabatic contraction can be well approximated by a model with spherical symmetry, homologous contraction, and conservation of  angular momentum.
This approximation assumes that all dark matter particles are on circular orbits such that $rM_{\rm enc}(r)$ (where $M_{\rm enc}$ is the total mass enclosed at radius $r$) is conserved upon infall.

In reality, dark matter can be on  highly eccentric orbits~\cite[see, e.g.,][]{Ghinga_1998}, and $rM_{\rm enc}(r)$ is not an adiabatic invariant. 
\cite{Gnedin_2004} therefore extends the \cite{Blumenthal_1986} model by showing that a better proxy for an invariant is $rM_{\rm enc}(\bar{r})$, where 
\begin{equation}
    \label{eqn:Gnedin_AC}
    \bar{r} = A \cdot R_{200} \left(\frac{r}{R_{200}}\right)^w
\end{equation}
is the {\it orbit-averaged} radius for elliptical orbits, $A$ is a normalization constant, and $w$ is a slope (in log space) parameter that together control how the instantaneous orbital radius $r$ is mapped onto its orbit-averaged value.
These fit parameters are determined empirically on a halo-by-halo basis~(see discussion in Appendix~\ref{appendix:Aandw}).

The conservation of radial action can be written as 
\begin{align}
\label{eqn:conservation}
\begin{split}
    &r_i\bigg[M_{{\rm DM}, i}(\bar{r}_i) + M_{{\rm baryon}, i}(\bar{r}_i) \bigg] \\
    &= r_f\bigg[M_{{\rm DM}, f}(\bar{r}_f) + M_{{\rm baryon}, f}(\bar{r}_f)\bigg]~,
\end{split}
\end{align}
where $M_{\rm DM}$ is the enclosed dark matter mass and $M_{\rm baryon}$ is the enclosed baryonic mass at radius $r$, or average radius $\bar{r}$,\ignorespaces
\footnote{\ignorespaces
Note that $r M(\bar{r})$ is a mixed combination, as it is the product of the radius $r$ and the mass enclosed within the orbit-averaged radius $\bar{r}$.
Using a set of high-resolution collisionless
simulations, \cite{Gnedin_2004} showed that this combination is better conserved than $\bar{r}M(\bar{r})$.
} 
in the initial~(subscript $i$) or final~(subscript $f$) state of the halo~\citep{Gnedin_2004,Hussein_2025}.
The goal of this subsection is to create a mapping from $M_{{\rm DM}, i}$ to $M_{{\rm DM}, f}$.
We take the ``initial'' state of the system as the $z=0$ DMO simulation and the ``final'' state as the $z=0$ hydro simulation.
In this way, $M_{{\rm DM},f}$ is exactly analogous to the mass-enclosed profile of the dark matter in the hydro simulation~($M_{\rm DM}^{\rm Hydro}$).
We therefore predict $M_{\rm DM}^{\rm Hydro}$ given only the simplistic conservation of the action in Equation~\ref{eqn:conservation}, referring to the result as $M_{\rm DM}^{\rm AC}$.

In practice, solving Equation~\ref{eqn:conservation} for $M_{\rm DM}^{\rm AC}$ requires a few additional assumptions.
For the term $M_{{\rm baryon},i}$, we assume that the (fictitious) ``baryonic component'' of a DMO simulation follows a self-similar mass-enclosed profile to the dark matter as a proxy for an uncollapsed halo~\citep[following from][]{Blumenthal_1986,Gnedin_2004,Binney_Tremaine_2008,Hussein_2025}.
Then, we numerically solve Equation~\ref{eqn:conservation} using a fixed-point method.
For each radius $r_i$, we solve for the corresponding contracted radius $r_f$ by rearranging Equation~\ref{eqn:conservation} to obtain
\begin{equation}
    \label{eqn:fixed_point}
    r_f = r_i \left(\frac{M_{\rm DM}^{\rm DMO}(\bar{r}_i)\cdot f_{n}(1-f_b) + M_{\rm DM}^{\rm DMO}(\bar{r}_i)\cdot f_{n}f_b}{M_{\rm DM}^{\rm DMO}(\bar{r}_i)\cdot f_{n}(1-f_b) + M_{\rm baryon}^{\rm Hydro}(\bar{r}_f)}\right)~\, ,
\end{equation}
where the `DMO'/`Hydro' superscript corresponds to the particular simulation that the enclosed mass pertains to.\ignorespaces
\footnote{\ignorespaces
Recall from Section~\ref{subsec:DREAMS} that the DMO simulations include an $\Omega_b$ fraction of baryons, which are treated as collisionless particles. This additional collisionless mass effectively increases the total dark matter mass in the DMO simulations by $\Omega_{b}$.
To account for this ``extra'' mass, we make the following correction:
\begin{equation}
    M_{\rm DM}^{\rm DMO,\,corrected} = M_{\rm DM}^{\rm DMO} \times \left(\frac{\Omega_{\rm M} - \Omega_b}{\Omega_{\rm M}}\right)~\, ,
\end{equation}
where $M_{\rm DM}^{\rm DMO}$ is the mass-enclosed profile for the DMO simulation.
From this point onwards, we refer to the ``corrected'' quantity when referencing $M_{\rm DM}^{\rm DMO}$.
}
Here, the total enclosed mass of the DMO halo is rescaled by 
\begin{equation}
    \label{eqn:fnorm}
    f_n = \frac{M_{\rm DM}^{\rm Hydro}(R_{200}) + M_{\rm baryon}^{\rm Hydro}(R_{200})}{M_{\rm DM}^{\rm DMO}(R_{200})}
\end{equation}
to ensure mass conservation with its hydro counterpart.  The fraction of baryons is also obtained from the hydro simulation:
\begin{equation}
    f_b = \frac{M_{\rm baryon}^{\rm Hydro}(R_{200})}{M_{\rm DM}^{\rm Hydro}(R_{200})} \, .
\end{equation}
By rescaling the dark matter mass profile in the DMO simulation~($M_{{\rm DM}}^{\rm DMO}$) by $f_{n}$ and $f_b$, we approximate the initial ``baryon'' component in the DMO simulation.\ignorespaces

\begin{figure*}
    \centering
    \includegraphics[width=0.98\linewidth]{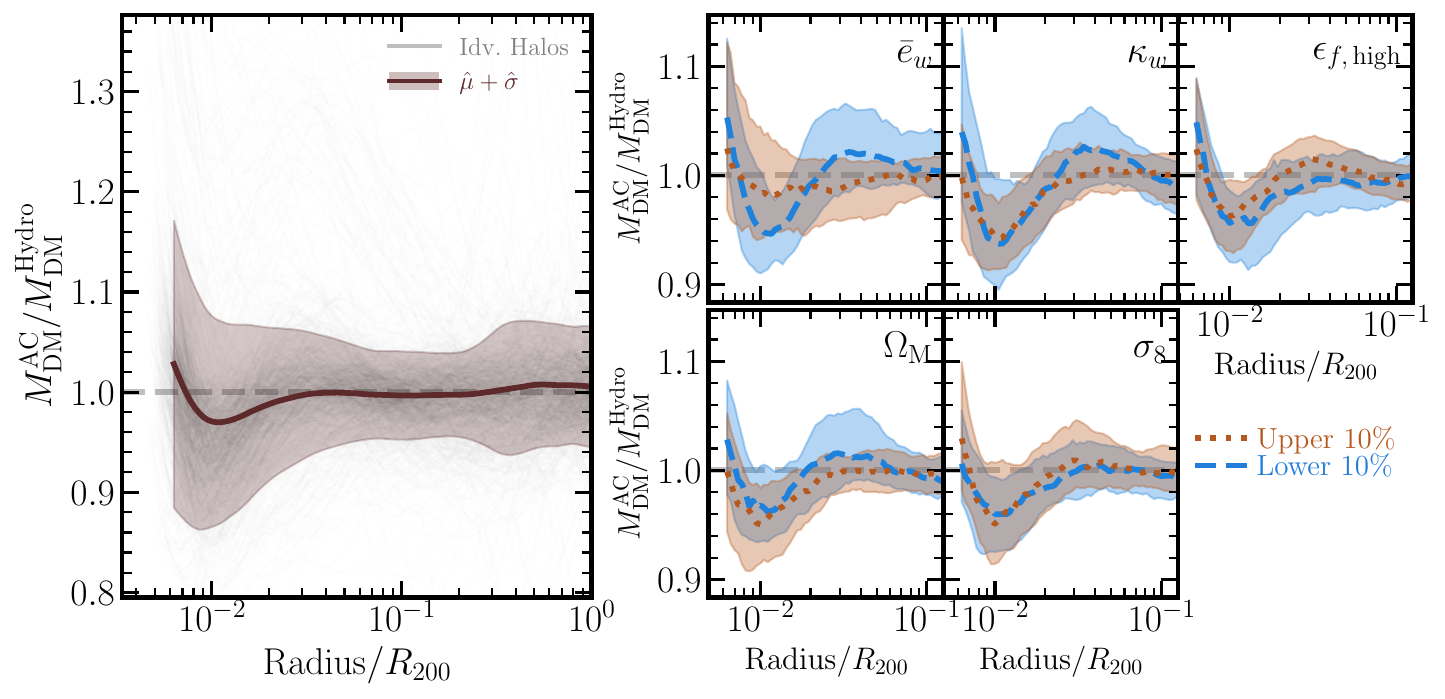}
    \caption{{\bf Ratio of Estimated Mass Contraction to Hydrodynamic Simulation Mass Profiles.}
    The ratio of the predicted adiabatically contracted dark matter mass profile ($M_{\rm DM}^{\rm AC}$; using the methods outlined in \protect\citeauthor{Gnedin_2004} \protect\citeyear{Gnedin_2004}) to the dark matter mass profile of the full hydrodynamic simulations ($M_{\rm DM}^{\rm Hydro}$), shown as a function of radius (rescaled by $R_{200}$.  Individual galaxies are shown by the gray lines.  
    The solid line represents the weighted mean of the distribution, while the shaded region is the weighted spread (see Section~\ref{subsec:weighting}).  The right-hand panels show the ratio for the same parameter ranges as the right-hand panels of Figure~\ref{fig:density_profiles}.  
    The level to which the halos agree with the adiabatically contracting halo model is mostly independent of astrophysics or cosmology parameter variations at radii $\gtrsim0.03R_{200}$ (with the exception of $\bar{e}_w$).
    }
    \label{fig:adiabatic_contraction}
\end{figure*}

The denominator of Equation~\ref{eqn:fixed_point} depends on both the enclosed mass of the dark matter and baryon components in the hydro simulation.  
The latter is calculated directly at radius $\bar{r}_f$. 
The former is set by assuming 
\begin{equation}
    \label{eqn:no_shell_crossing}
 M_{\rm DM}^{\rm AC}(\bar{r}_f) =  M_{\rm DM}^{\rm DMO}(\bar{r}_i) \, .
\end{equation}
Equation~\ref{eqn:no_shell_crossing} ensures that the contraction proceeds in a strictly monotonic fashion, i.e., that shells of dark matter preserve their radial ordering and do not cross one another.

Equation~\ref{eqn:fixed_point} provides a map that relates each radius in the DMO simulation to its contracted counterpart in the hydrodynamic run.
With this mapping from $r_i$ to $r_f$, we evaluate $M_{\rm DM}^{\rm DMO}$ at $r_f$ to determine our prediction for $M_{\rm DM}^{\rm AC}$.

The main panel of Figure~\ref{fig:adiabatic_contraction} shows the predictions for $M_{\rm DM}^{\rm AC}$, normalized by the true $M_{\rm DM}^{\rm Hydro}$, as a function of radius.
From $\sim0.03R_{200}$ to $R_{200}$, the predicted adiabatically contracted profile is largely consistent with the hydro simulation.
At smaller radii~($\lesssim0.02R_{200}$), however, the deviations are more notable, likely because this region is near the center of the disk where the baryon modeling becomes more important.
Here, the effects of feedback are likely to be stronger due to, e.g., the higher concentration of supernovae relative to larger radii.
Specifically, there is a significant under-prediction of the mass enclosed at $\sim0.01R_{200}$ and a subsequent over-prediction at radii $<0.01R_{200}$, although there is significant scatter.
We attribute the deviations at these small radii to the baryon feedback preventing the halo from contracting, or doing so adiabatically.

Each of the panels on the right of Figure~\ref{fig:adiabatic_contraction} show the mass-contraction ratio broken down by feedback and cosmology variations. The extent of the deviations from the adiabatic contraction model depends on the strength of supernova feedback ($\bar{e}_w$), with stronger feedback corresponding to slightly better agreement with the \cite{Gnedin_2004} adiabatic contraction model.
However, the bulk of the deviations from the adiabatic contraction model, at all radii, are mostly on the order of a few percent.
Even in the most central regions, the predictions vary by $\sim3$--$10\%$ from the hydro simulations.
Overall, the agreement between predictions and simulation results suggests that the \cite{Gnedin_2004} approximation for an adiabatically contracting halo with elliptical orbits is reasonable for the DREAMS halos.

A small fraction of halos~($27$~\edit{out of the 1024; $2.6\%$}) deviate significantly~($\gg10\%$) from the adiabatic contraction model at $R_{200}$.
The large deviation at this radius is surprising considering that part of the calculation above is to renormalize the DMO profile based on masses at the virial radius~(e.g., Equation~\ref{eqn:fnorm}).
Upon further inspection, all of the cases where the profiles deviate significantly have a merger/large accretion event crossing the virial radius at $z\sim0$.
The exact timing of this event is different between the DMO and hydrodynamic simulations~(which could be for a number of reasons, physical or numerical; \citeauthor{Genel_2019} \citeyear{Genel_2019}, \citeauthor{Pakmor_2025} \citeyear{Pakmor_2025}).
The required renormalization of the DMO mass profile can therefore be systematically over- or under-estimated in situations where the infalling material crosses the virial radius at different times, leading to a failure of the adiabatically contracting halo model.
Indeed, a halo undergoing a merger/large accretion event is likely to not be adiabatically~(or even quasi-adiabatically) contracting \citep{Velmani_2023}.

\subsubsection{Comparison with Bursty Feedback Model}
\label{subsec:bursty}

\begin{figure}
    \centering
    \includegraphics[width=0.98\linewidth]{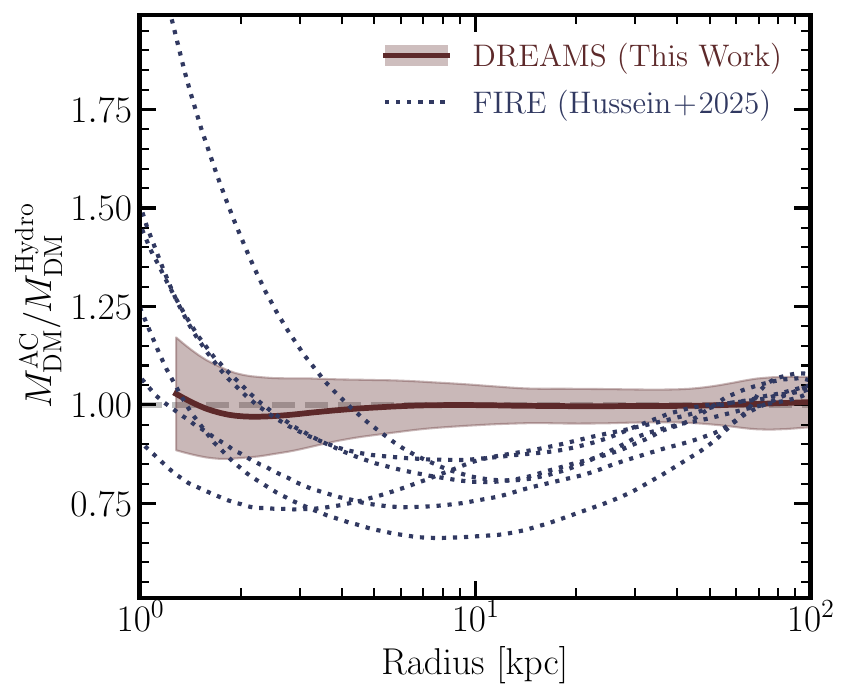}
    \caption{{\bf Role of Bursty Feedback in Adiabatic Contraction.}
    The ratio between the DM mass predicted using adiabtic contraction \protect\citep[using][]{Gnedin_2004} and the DM mass in hydro simulations.
    The results are shown here for both FIRE-2 (data from \protect\citeauthor{Hussein_2025} \protect\citeyear{Hussein_2025}; note that each line is an individual halo) and DREAMS (same as Figure~\ref{fig:adiabatic_contraction} without radial normalization; solid line is mean, shaded region is standard deviation).
    The {\it bursty} feedback model, FIRE, varies far more significantly than systematic feedback variations within DREAMS~(see, e.g., right-hand panels of Figure~\ref{fig:adiabatic_contraction}).
    This suggests that deviations from adiabatic contraction are driven by {\it bursty} feedback and not just {\it strong} feedback.
    }
    \label{fig:comp_to_Abdelaziz}
\end{figure}

Virtually all of the DREAMS simulated halos agree with a simple adiabatically contracting model within $\sim10\%$ at all radii~(Figure~\ref{fig:adiabatic_contraction}).
This is in qualitative agreement with the findings of \citeauthor{Hussein_2025}~(\citeyear{Hussein_2025}) who use the TNG \citep{Pillepich_2018}, AURIGA \citep{Grand_2017}, and VINTERGATAN \citep{Agertz_2021,Rey_2023} models.
TNG and AURIGA model the star-forming ISM with an effective equation of state that results in smooth (i.e., non-bursty) stellar feedback (\citeauthor{Springel_Hernquist_2003}~\citeyear{Springel_Hernquist_2003}).
VINTERGATAN, on the other hand, does have an explicit ISM model like FIRE, but VINTERGATAN does not allow stars to form at 100\% efficiency per freefall time as FIRE does (instead using 10\%; \citeauthor{Agertz_2013}~\citeyear{Agertz_2013}).
The decrease in the star formation efficiency leads to a decrease in the total feedback energy of supernovae, leading to less severe blowouts of gas in VINTERGATAN compared to FIRE. 

\cite{Hussein_2025} also made comparisons to the FIRE-2 model \citep{FIRE2} and found that the FIRE model varies significantly (factors of $\gtrsim2$; shown in Figure~\ref{fig:comp_to_Abdelaziz}) from the simple case of an adiabatically contracting halo.
Figure~\ref{fig:comp_to_Abdelaziz} demonstrates that the element of the FIRE model that is causing the deviations from adiabatic contraction is the {\it bursty} feedback---not just strong feedback.
As we showed in the previous subsection, the feedback variations in the TNG model are sufficiently strong to almost entirely prevent the build up of a stellar component in the galaxy; however, the TNG model does not have bursty feedback.
Variations from an adiabatically contracting halo in even the most extreme TNG feedback variations are significantly less than those of FIRE-2, suggesting that bursty, episodic feedback plays the strongest role setting deviations from the adiabatically contracting halo.

It is not clear, however, whether bursty feedback disrupting the potential and preventing adiabatic contraction is a generic feature of FIRE-2 or a feature of the specific implementation of the model.
An analogous suite of Milky Way-mass halos within the FIRE-2 model (or its successor FIRE-3; \citeauthor{FIRE3} \citeyear{FIRE3}) would be particularly valuable for constraining this.
Such a comparison would help determine whether the deviations from \edit{simple} adiabatic contraction \edit{models} identified by \cite{Hussein_2025} are a generic outcome of bursty feedback or depend sensitively on the details of how feedback is modeled in FIRE.

\section{Conclusion} \label{sec:conclusions}

In this work, we investigated the impact of variations in the IllustrisTNG physical model and cosmology on the dark matter density profiles of Milky Way-mass halos using the DREAMS simulations.
The DREAMS CDM suite consists of 1024 hydrodynamic simulations with variations in supernova feedback, AGN feedback, and cosmology~(see, e.g., Figure~\ref{fig:pretty_picture} for a few such halos), as well as a corresponding set of 1024 DMO simulations.
These simulation suites enabled us to quantify the role that baryons play in the assembly of dark matter halos.

Our conclusions are as follows:
\begin{itemize}[leftmargin=10pt]
    \item Overall, there are only minor variations in the density profiles of Milky Way-mass halos with variations to the IllustrisTNG model~(Section~\ref{subsec:density_profiles} Figure~\ref{fig:density_profiles}). 
    Virtually all variation in the density profiles is driven by halo-to-halo variation, with a relatively small contribution from the two supernova parameters that only create differences of $\pm \sim20-50\%$ in the inner regions ($\lesssim 0.01R_{200}$) of the halos.
    
    \item We fit the density profiles with a generalized NFW profile~\citep[e.g.,][]{Jaffe_1983,Hernquist_1990} which has two normalization parameters and three shape parameters~(Equation~\ref{eqn:gNFW}, Section~\ref{subsec:gNFW}).
    Again, intrinsic halo-to-halo variation plays the dominant role in setting both the normalization and shape of halos.
    Beyond the halo-to-halo variance, changes to the supernova wind speed ($\kappa_w$) can also vary the normalization of the profile (Figure~\ref{fig:gNFW_norm}) and increases to the wind energies ($\bar{e}_w$) can \edit{reduce the deviations from an} NFW \edit{profile} (Figure~\ref{fig:gNFW}).
    Physically, this suggests that ejecting {\it faster} winds more strongly disrupts the potential, whereas {\it more} wind energy effectively suppresses stellar mass growth.

    \item \edit{We quantify the impact of baryons on the dark matter halos by defining the central mass growth as ratio of the DMO simulation enclosed mass to the hydro simulation enclosed mass at $0.01R_{200}$~(Figure~\ref{fig:mass_growth}).}
    Here, the astrophysics variations play a dominant role.
    Larger values of $\bar{e}_w$ and $\kappa_w$ are more efficient at preventing stellar mass growth~(left two panels of Figure~\ref{fig:stellar_mass}), thus the halo contracts less.
    On the other hand, we find the opposite trend in $\epsilon_{f,\,\mathrm{high}}$: increasing black hole feedback increases the work required to contract~(right panel of Figure~\ref{fig:stellar_mass}, see also Appendix~\ref{appendix:AGN_anamoly}).
    
    \item We compared the results of the hydro simulations to DMO simulations with matched initial conditions~(Section~\ref{subsec:dmo}).
    On the whole, the DMO simulations have a lower density in the inner regions ($\lesssim0.01R_{200}$) than the hydro simulations~(Figure~\ref{fig:comp_to_DMO}).
    The extent to which the density is enhanced depends on the feedback, however.
    In fact, higher values of $\bar{e}_w$ tend towards a ratio of unity with respect to the DMO simulations, in good agreement with the results of the profile shape and central mass growth.

    \item We quantified whether or not our halos are consistent with approximations to adiabatic contraction~(Section~\ref{subsec:adiabatic_contraction}).
    Halos generally agree within $\sim10\%$ regardless of the feedback or cosmology~(Figure~\ref{fig:adiabatic_contraction}); however, systems with late-time mergers that occur at slightly different times between the hydro and DMO simulations tend to disagree significantly.
    As with many previous results, the supernova wind energy changes the picture due to the lack of baryon mass accumulation: these halos contract adiabatically.
    Finally, we showed that the bursty feedback FIRE(-2) model produces significantly larger deviations from adiabatic contraction than the average DREAMS halos~(Figure~\ref{fig:comp_to_Abdelaziz}).
    This suggests that what drives deviations from adiabatic contractions are {\it bursts} of stellar feedback, not just strong feedback.
\end{itemize}

Taken together, these results show that halo-to-halo variance is the dominant source of scatter in Milky Way–mass dark matter profiles in the TNG model~\edit{(within the DREAMS parameter variations)}, with $\bar{e}_w$ and $\kappa_w$ feedback parameters producing secondary but non-negligible effects.
AGN feedback plays a smaller role, and cosmological parameter variations are essentially irrelevant at this mass scale.

Understanding and quantifying the range of predictions in an individual simulation model is a difficult task, yet it is necessary to fully understand that model.
This work, which is part of a broader series in the DREAMS Project, represents one such investigation into the sensitivity of the TNG model to its input physics within a $\Lambda$CDM context.
Such a benchmark is useful in the wider landscape of future simulation efforts to vary different simulation models, feedback physics, and dark matter physics within the DREAMS collaboration.

\section*{Data Availability}

Raw data from the DREAMS CDM suite is available via Globus transfer at \href{https://dreams-project.readthedocs.io/en/latest/data_access.html}{dreams-project.readthedocs.io/en/latest/data\_access.html}.
The scripts used to generate the reduced data and all other scripts used to support the findings of this work are made publicly available \dataset[doi: 10.5281/zenodo.18715476]{\doi{10.5281/zenodo.18715476}} and on github \href{https://github.com/AlexGarcia623/DREAMS-Density-Profiles/tree/main}{github.com/AlexGarcia623/DREAMS-Density-Profiles}

\section*{Acknowledgements}

The authors gratefully acknowledge the use of computational resources and support provided by the Scientific Computing Core at the Flatiron Institute, a division of the Simons Foundation.
The authors also acknowledge Research Computing at The University of Virginia for providing computational resources and technical support that have contributed to the results reported within this publication  (URL: \href{https://rc.virginia.edu}{https://rc.virginia.edu}), as well as Princeton University's Research Computing resources.

AMG acknowledges a helpful conversation with Sebastian Trujillo-Gomez as a part of the Simulation Based Inference Conference.
AMG thanks Carrie Filion for helpful comments that led to the success of this work.
Finally, AMG acknowledges Kate Garcia for assistance in the design of Figure~\ref{fig:pretty_picture} and Kelly Garcia for assistance in the design of Figure~\ref{fig:gNFW}.

AMG, PT, JK, NA, AB, XO, AF, and NK acknowledge support from the National Science Foundation under Cooperative Agreement 2421782 and the Simons Foundation grant MPS-AI-00010515 awarded to NSF-Simons AI Institute for Cosmic Origins -- CosmicAI, \href{https://www.cosmicai.org/}{https://www.cosmicai.org/}.
ML is supported by the Department of Energy~(DOE) under Award Number DE-SC0007968, as well as the Simons Investigator Award. AC acknowledges the hospitality of the Weizmann Institute of Science and the support from the Benoziyo Endowment Fund for the Advancement of Science. This project has received funding from the European Research Council (ERC) under the European Union’s Horizon Europe research and innovation programme (grant agreement No.
101117510). Views and opinions expressed are however
those of the authors only and do not necessarily reflect those of the European Union. The European Union cannot be held responsible for them.
LN is supported by the Sloan Fellowship, the NSF CAREER award 2337864, NSF award 2307788, and by the NSF award PHY2019786 (The NSF AI Institute for Artificial Intelligence and Fundamental Interactions, \href{http://iaifi.org/}{http://iaifi.org/})
XS acknowledges the support of the NASA theory grant JWST-AR-04814. This work was performed in part at Aspen Center for Physics, which is supported by National Science Foundation grant PHY-2210452. The authors also thank the Simons Foundation for their support in hosting and organizing workshops on the DREAMS Project.  
HL is supported by the U.S. Department of Energy under grant DE-SC0026297 and the Cecile K. Dalton Career Development Professorship, endowed by Boston University trustee Nathaniel Dalton and Amy Gottleib Dalton.

\appendix

\section{Emulator Validation Test}
\label{appendix:emulator_validation}

\setcounter{equation}{0}
\setcounter{figure}{0} 
\setcounter{table}{0}
\renewcommand{\theequation}{A\arabic{equation}}
\renewcommand{\thefigure}{A\arabic{figure}}
\renewcommand{\thetable}{A\arabic{table}}

\begin{figure*}
    \centering
    \includegraphics[width=0.9\linewidth]{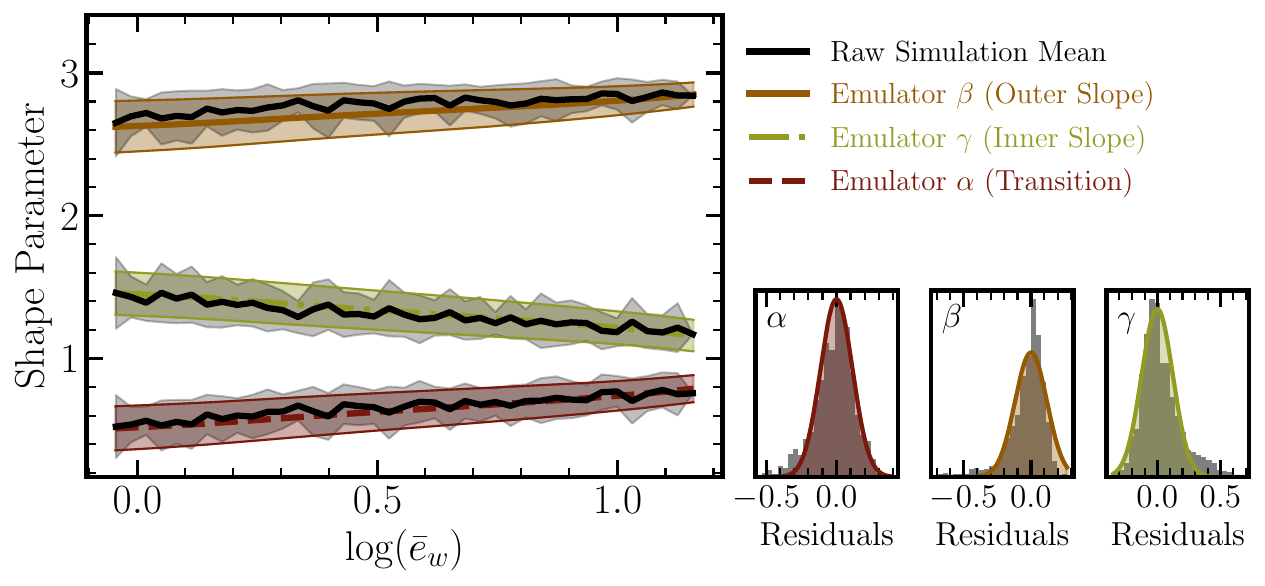}
    \caption{{\bf Example Emulator Validation.}
    {\it Left:} Same as Figure~\ref{fig:gNFW}, with the raw estimate of the rolling mean and standard deviation (black solid line and gray shaded region) for each parameter.
    {\it Right:} Histograms of the residuals about the true median for $\alpha$ (left), $\beta$ (middle), and $\gamma$ (right) along with a Gaussian distribution consistent with the average standard deviation predicted by the emulator.
    }
    \label{fig:appendix_emulator_validation}
\end{figure*}

This work employs a series of neural network emulators to coherently interpret the DREAMS data.
We validate each emulator (Figures~\ref{fig:gNFW_norm},~\ref{fig:gNFW},~\ref{fig:mass_growth},~\ref{fig:stellar_mass},~\ref{fig:appendix_bh_mass},~and~\ref{fig:appendix_AC2}) by comparing its predictions to the raw simulation outputs~(following \citetalias{Rose_2025}).
Figure~\ref{fig:appendix_emulator_validation} shows an example of this validation process.
The left panel of Figure~\ref{fig:appendix_emulator_validation} is identical to Figure~\ref{fig:gNFW} with the addition of the true rolling mean (solid black line) and standard deviation (gray shaded region) from the measured simulation data points.  The raw simulation results largely track the emulator ones, although there is a small shift in the outer slope. 
One potential reason for this is that the emulator is trained to predict single parameter dependencies (i.e., we keep all other parameters fixed), whereas the raw simulation outputs include inherent co-variation of parameters.
To capture this co-variance, we would need a different architecture, such as normalizing flows \citep[see, e.g.,][]{CDM_Velocity}.
Regardless, the emulated prediction for the mean is largely consistent with these raw simulation outputs even with our simple emulator. 
Figure~\ref{fig:appendix_emulator_validation} further demonstrates the utility of the emulator: the true rolling mean is significantly noisier than the predictions from the emulated sample.
The right-hand panels of Figure~\ref{fig:appendix_emulator_validation} show the distribution of raw residuals compared to an average estimate from the emulator.\ignorespaces
\footnote{\ignorespaces
The scatter about the shape parameters can vary as a function of $\bar{e}_w$ in both the raw dataset and emulator; here, we only compare the full distribution of residuals against the average of the standard deviations from the emulator.}
There is some level of disagreement in the tails of the distributions (in particular in $\gamma$), which, again, could be more effectively captured by a more sophisticated architecture, but for the most part we find that the standard deviation of the datasets match quite well with the raw simulation data.

We perform the same test for each of the emulators built in this work and find a similar level agreement.

\section{Assessing the Role of \texorpdfstring{$\epsilon_{f,\,\mathrm{high}}$}{A AGN}}
\label{appendix:AGN_anamoly}

\setcounter{equation}{0}
\setcounter{figure}{0} 
\setcounter{table}{0}
\renewcommand{\theequation}{C\arabic{equation}}
\renewcommand{\thefigure}{C\arabic{figure}}
\renewcommand{\thetable}{C\arabic{table}}

\begin{figure}
    \centering
    \includegraphics[clip,trim={0 0 0 11.07cm},width=\linewidth]{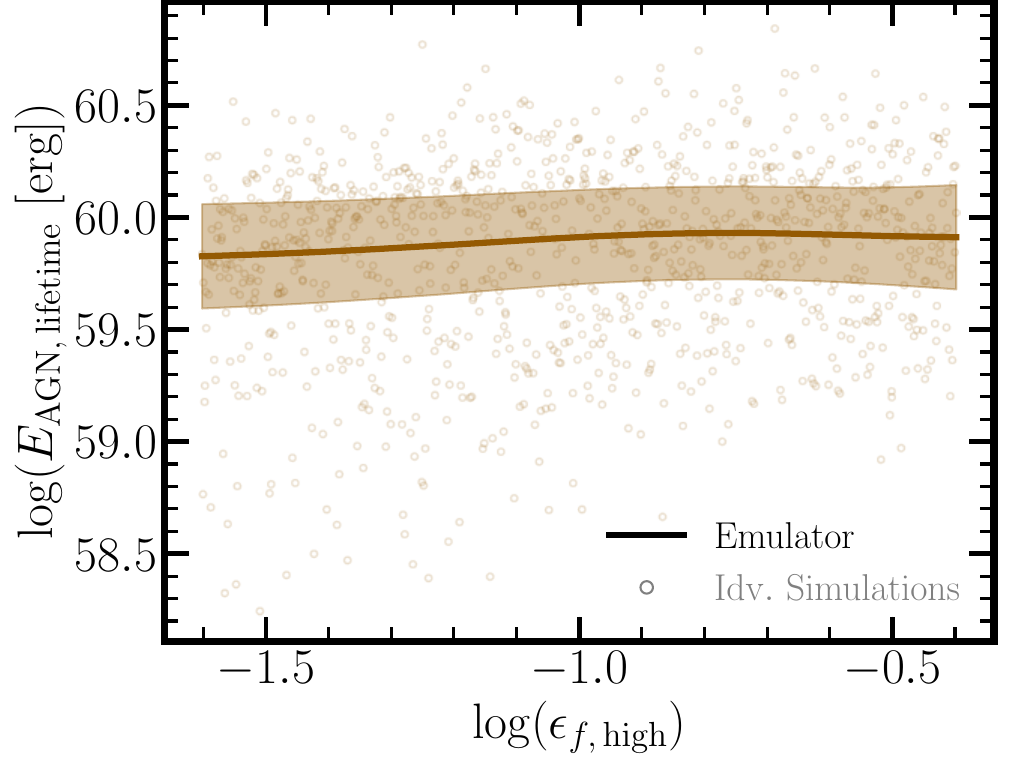}
    \caption{{\bf Approximate AGN Feedback Budget.}
    The approximated lifetime feedback energy of the AGN (via Equation~\ref{eqn:agn_lifetime_feedback}) in each simulation (open circles).
    The solid line, and shaded region thereof, represents the predictions from an ensemble of emulators for the mean and standard deviation, respectively.
    }
    \label{fig:appendix_bh_mass}
\end{figure}

There are notable trends between physical properties of the halo and the AGN feedback.
Section~\ref{subsubsec:mass_growth} lays out an argument for these trends based on an approximation of the total feedback energy~(by integrating Equation~\ref{eqn:agn}; see also a similar argument made by \citeauthor{Ni_2023}~\citeyear{Ni_2023}).
This appendix provides supplementary support for this interpretation. 

Figure~\ref{fig:appendix_bh_mass} shows 
our approximation for the total feedback energy imparted by the central black hole~(via Equation~\ref{eqn:agn_lifetime_feedback}; open circles).
In addition, we train a series of emulators~(see Section~\ref{subsec:emulator}) and report the aggregate prediction for the mean and standard deviation as the solid line and shaded region, respectively.
The total feedback energy over the lifetime of the black hole is virtually unchanged $(\sim 10^{60}~{\rm erg})$ by variations in $\epsilon_{f,\,\mathrm{high}}$.
It decreases {\it slightly} for feedback values less than the fiducial TNG $(\log(\epsilon_{f,\,\mathrm{high}}) \lesssim-1)$, but it is completely flat above this value.
This suggests that the total work that the AGN does (in the thermal mode) is roughly constant as a function of $\epsilon_{f,\,\mathrm{high}}$.

We can write a toy model for the balance of feedback from the AGN with the accretion onto the black hole by equating the energy via feedback to the binding energy of the surrounding gas. For a system in 
equilibrium, we can approximate Equation~\ref{eqn:agn_lifetime_feedback} with the gravitational binding energy of the gas such that
\begin{equation}
    \label{eqn:binding_energy}
    \epsilon_{f,\,\mathrm{high}}\epsilon_r M_{\rm BH}c^2 \approx G \frac{M_{\rm gas}^2}{r}~,
\end{equation}
where $M_{\rm gas}$ is the mass of gas within a radius $r$ of the black hole, and $G$ is the gravitational constant.
We can then rearrange Equation~\ref{eqn:binding_energy} to find $M_{\rm BH}\propto \epsilon_{f,\,\mathrm{high}}^{-1}$.
While a crude and incomplete physical picture of the complexities of the center of galaxies~(e.g., the $M_{\rm gas}$ term may also have complex dependencies on AGN feedback variations), it is interesting to note that the inverse scaling of $M_{\rm BH}$ with increasing $\epsilon_{f,\,\mathrm{high}}$ is in broad agreement with the trends we find in the DREAMS simulations~(Figure~\ref{fig:appendix_bh_mass}).

\section{Adiabatic Contraction Fit Parameters}
\label{appendix:Aandw}

\setcounter{equation}{0}
\setcounter{figure}{0} 
\setcounter{table}{0}
\renewcommand{\theequation}{D\arabic{equation}}
\renewcommand{\thefigure}{D\arabic{figure}}
\renewcommand{\thetable}{D\arabic{table}}

\begin{figure*}
    \centering
    \includegraphics[width=0.48\linewidth]{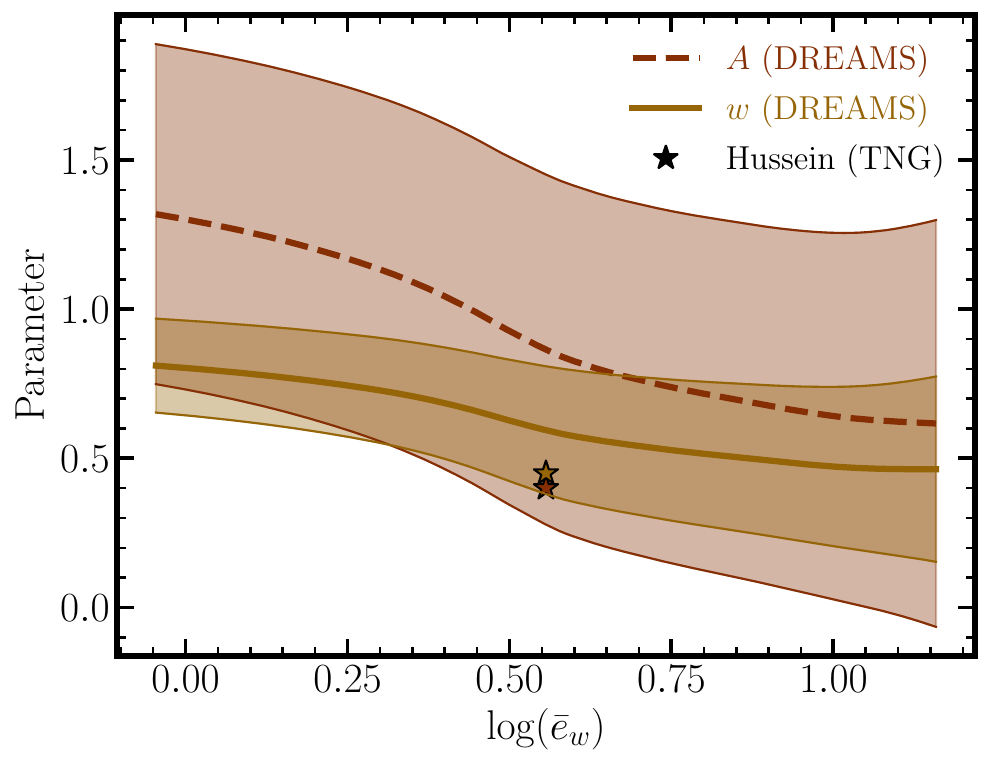}
    \includegraphics[width=0.48\linewidth]{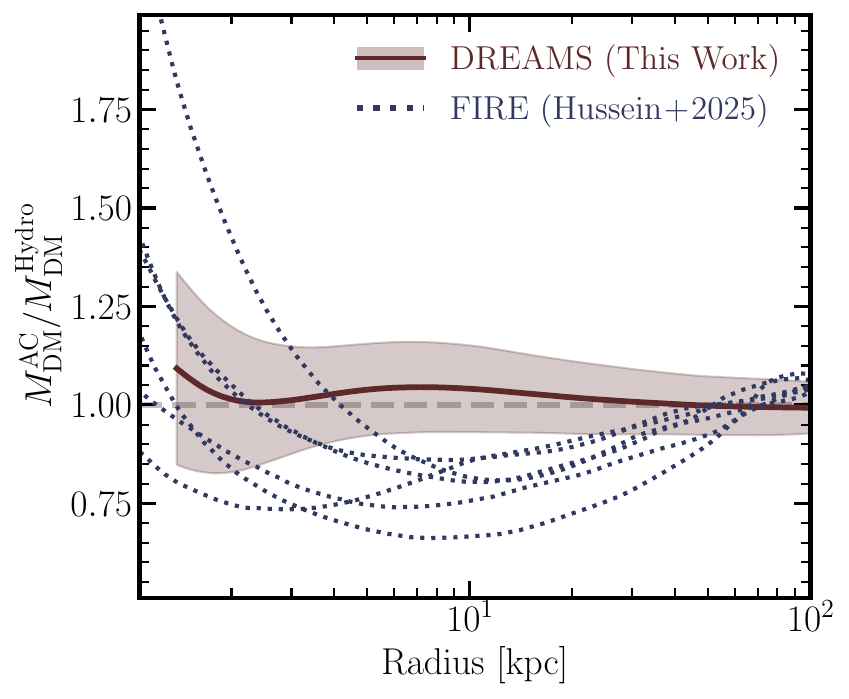}
    \caption{{\bf Dependence of $\mathbf{A}$ and $\mathbf{w}$ on DREAMS feedback variations.}
    {\it Left:} Predictions from our emulators for the best-fit $A$ and $w$ values~(from Equation~\ref{eqn:Gnedin_AC}) as a function of $\bar{e}_w$ ($A$ and $w$ do not depend sensitively on any other DREAMS simulation parameter).
    The solid line shows the mean while the shaded region represents the halo-to-halo variations.
    The stars correspond to the average values from \protect\citeauthor{Hussein_2025}~(\protect\citeyear{Hussein_2025}; $A=0.40$ and $w=0.45$), which fall well within the halo-to-halo variation in DREAMS.
    {\it Right:}  Same as Figure~\ref{fig:comp_to_Abdelaziz}, but assuming a fixed $A$ and $w$ value that corresponds to the average of the DREAMS distribution ($A=0.848$ and $w=0.675$).
    We still find qualitatively different behavior between FIRE-2 and DREAMS.
    }
    \label{fig:appendix_AC2}
\end{figure*}

As mentioned in Section~\ref{subsec:adiabatic_contraction}, the \cite{Gnedin_2004} adiabatic contraction model includes a best-fit parameterization of the orbit-averaged radius $\bar{r}$~(Equation~\ref{eqn:Gnedin_AC}).
In our work, the free parameters, $A$ and $w$ (corresponding to a normalization and exponent of the fit), are determined on a halo-by-halo basis.
Not every work follows this practice; 
for example, \cite{Hussein_2025} fit their halos using the {\it average} $A$ and $w$ parameters.
We opt against this because $A$ and $w$ depend on the physics variations that DREAMS employs.
In particular, they both depend on $\bar{e}_w$ (left panel of Figure~\ref{fig:appendix_AC2}).
With increasing $\bar{e}_w$, both $A$ and $w$ decrease (though the former decreases more than the latter).
These trends follow fairly closely to those observed with the central mass growth in Figure~\ref{fig:mass_growth}, i.e., halos that undergo more contraction have larger $A$ and $w$ values, which physically correspond to systematically smaller orbit-averaged radii, indicating that dark matter has moved deeper into the potential well.

If we instead use the average values of $A=0.848$ and $w=0.675$, our results are qualitatively unchanged.\ignorespaces
\footnote{\ignorespaces
These average values of $A$ and $w$ are different than those reported in \citeauthor{Hussein_2025}~(\citeyear{Hussein_2025}; $A=0.40$ and $w=0.45$), who use six halos in the TNG model (albeit without the physics variations).
However, we find that the \cite{Hussein_2025} values are consistent with our results within the halo-to-halo variation (shown as the stars in the left panel of Figure~\ref{fig:appendix_AC2}).
}
The right-hand panel of Figure~\ref{fig:appendix_AC2} shows the resulting adiabatic contraction profiles when fixing $A$ and $w$.
On average, there is slightly worse agreement between the adiabatic contraction model and the true hydrodynamic simulations, but the level to which the two disagree is within $\sim10\%$.
The (weighted) scatter also increases when considering fixed $A$ and $w$.
Our key conclusion regarding the comparison with the bursty FIRE feedback model remains qualitatively unchanged, however.
Even when ignoring the variations of $A$ and $w$ due to the physics variations,
we find that the deviations seen in FIRE halos are significantly larger than those of the DREAMS halos.

\bibliography{paper}{}
\bibliographystyle{aasjournal}
\end{document}